\documentclass[12pt]{iopart}

\usepackage{graphicx}
\usepackage{dcolumn}
\usepackage{bm}
\usepackage{amssymb,amsfonts}
\usepackage{amsthm,mathbbol}
\usepackage[utf8]{inputenc}
\usepackage[T1]{fontenc}
\usepackage{mathptmx}
\usepackage{enumerate}   
\usepackage{hyperref}   
\usepackage{soul}
\usepackage[normalem]{ulem}
\DeclareSymbolFont{greekletters}{OML}{cmr}{m}{it}
\DeclareMathSymbol{\varrho}{\mathord}{greekletters}{"25}

\usepackage{cite}
\usepackage{xcolor}

\begin{document}

\title[ Classical and Quantum Approaches to Out-of-Equilibrium Statistical Mechanics]{Boltzmann to Lindblad: Classical and Quantum Approaches to Out-of-Equilibrium Statistical Mechanics}

\author{Stefano Giordano}%
\address{University of Lille, CNRS, Centrale Lille, Univ. Polytechnique Hauts-de-France, UMR 8520 - IEMN - Institut d'{\'E}lectronique, de Micro{\'e}lectronique et de Nanotechnologie, F-59000 Lille, France}
\ead{stefano.giordano@univ-lille.fr}  

\author{Giuseppe Florio}%
\address{Department of Civil Environmental Land Building Engineering and Chemistry, DICATECh, Polytechnic University of Bari, via
Orabona 4, I-70125 Bari, Italy}
\address{INFN, Section of Bari, I-70126, Italy}
\ead{giuseppe.florio@poliba.it}  

\author{Giuseppe Puglisi}%
\address{Department of Civil Environmental Land Building Engineering and Chemistry, DICATECh, Polytechnic University of Bari, via
Orabona 4, I-70125 Bari, Italy}
\ead{giuseppe.puglisi@poliba.it}  

\author{Fabrizio Cleri}%
\address{University of Lille, Institut d'{\'E}lectronique, de Micro{\'e}lectronique et de Nanotechnologie (IEMN CNRS UMR8520) and Departement de Physique, F-59652 Villeneuve d'Ascq, France}
\ead{fabrizio.cleri@univ-lille.fr}  

\author{Ralf Blossey}
\address{University of Lille, Unit{\'e} de Glycobiologie Structurale et Fonctionnelle (UGSF), CNRS UMR8576, F-59000 Lille, France}
\ead{ralf.blossey@univ-lille.fr}

\begin{abstract}

Open quantum systems play a central role in contemporary nanoscale technologies, including molecular electronics, quantum heat engines, quantum computation and information processing. A major theoretical challenge is to construct dynamical models that are simultaneously consistent with classical thermodynamics and with the requirement of complete positivity of quantum evolution. In this work, we develop a framework that addresses this issue by systematically extending classical stochastic dynamics to the quantum domain. We begin by formulating a generalized Langevin equation in which both friction and noise act symmetrically on the two Hamiltonian equations. From this, we derive a generalized Klein–Kramers equation expressed in terms of Poisson brackets, and we show that it admits the Boltzmann distribution as its stationary solution while satisfying the first and second laws of thermodynamics along individual trajectories.

Applying canonical quantization to this classical framework yields two distinct quantum master equations, depending on whether the friction operators are taken to be Hermitian or non-Hermitian. By analyzing the dynamics of a harmonic oscillator, we determine the conditions under which these equations reduce to a Lindblad-type generator. Our results demonstrate that complete positivity is ensured only when friction and noise are included in both  Hamiltonian equations, thus fully justifying the classical construction. Moreover, we find that the friction coefficients must satisfy the same positivity condition in both the Hermitian and non-Hermitian formulations, revealing a form of universality that transcends the specific operator representation.

The formalism developed here provides a thermodynamically consistent and completely positive quantum extension of classical stochastic mechanics. It offers a versatile tool for deriving quantum versions of the thermodynamic laws and is directly applicable to a wide class of nonequilibrium nanoscale systems of current theoretical and technological interest.

\end{abstract}


\submitto{\JSTAT}


\newpage

\section{Introduction}
\label{intro}

The study of open quantum systems emerged from the recognition that quantum systems are rarely isolated, as their coupling to external environments inevitably induces noise, dissipation, and decoherence.
An early attempt to include these effects within quantum mechanics led to the Caldirola–Kanai equation \cite{caldirola1941,kanai1948}, derived by incorporating nonconservative forces into Schrödinger’s framework through a nonlinear time transformation originally proposed by Levi-Civita.
Although initially controversial because of this unusual transformation of time, this approach was later shown to be consistent with the physics of open systems and laid the foundation for quantum dissipation theory \cite{caldirola1982,caldirola1983}, which was subsequently developed in numerous works \cite{svinin1975,stocker1979,dekker1981}.

A major advance came with the formulation of the quantum master equation, describing the evolution of an open system’s density matrix \cite{redfield1965,blum1981,lindenberg1990,zwanzig2001}.
The Redfield equation, valid for weak system–environment coupling \cite{redfield1965}, found early applications in nuclear magnetic resonance but does not ensure positivity of the density matrix \cite{trushechkin2021}.
This limitation motivated the development of more rigorous approaches, including path-integral formulations, quantum thermodynamics, and quantum Brownian motion \cite{caldeira1981,caldeira1983,caldeira1983bis,beretta1984,beretta1985,jannussis1985,ohba1997,cavalcanti1998}.
In the 1970s, these efforts culminated in the Gorini–Kossakowski–Sudarshan–Lindblad (GKLS) equation \cite{lindblad1974,lindblad1975,gorini1976,lindblad1976,pascazio2017}, which provides a fully Markovian and mathematically consistent framework for quantum dynamical semigroups, refining the Redfield approximation \cite{redfield1965}, and advancing the theoretical foundations of quantum thermodynamics \cite{spohn1978,lebowitz1978,trushechkin2017,trushechkin2019,ruelle2001,pillet2002,kosloff2019,prior2010,campaioli2024}.
This framework clarified the mechanisms of decoherence and environmental coupling \cite{schlosshauer2004,toth2014,schlosshauer2019} and inspired subsequent studies on quantum equipartition \cite{bialas2018,spiechowicz2018,bialas2019,luczka2020,tong2024} and fluctuation theorems \cite{tal2,tal3}.

Today, open quantum system theory constitutes a cornerstone of modern physics.
It underpins applications in nuclear magnetic resonance \cite{abragam1961,slichter1990}, optical pumping \cite{happer1972}, masers \cite{scovil1959,thomas2020}, and lasers \cite{youssef2009,dorfman2018}, as well as in quantum computation \cite{deutsch1985,buluta2009,mermin2012,cleri2024}, quantum information \cite{vedral2006,hayashi2017}, and molecular electronics \cite{ke2008,jensen2022}.
The same framework informs quantum biology \cite{landsberg1984,mohseni2014,bianco1992,chiabrera2000}, notably in photosynthetic energy transfer \cite{engel2007,sarovar2010,scholes2011}, and plays a central role in quantum thermodynamic devices \cite{myers}, quantum batteries \cite{dutta2021}, heat engines \cite{nori2007}, and in understanding the thermodynamic arrow of time \cite{maddox1985,lebowitz1993,guff2025}.
Furthermore, it provides the theoretical foundation for quantum metrology and sensing \cite{giovannetti2011,pezze2018,braun2018}, optomechanics \cite{barzanjeh2022,huang2024}, nonequilibrium phenomena in nanoscale and condensed-matter systems \cite{keimer2017,tokura2017,alfieri2023,goyal2025}, and engineering of active quantum matter \cite{lowen}.
{  In many realistic physical situations, the interaction between a system and its environment gives rise to memory effects, making non-Markovian descriptions of open quantum dynamics both necessary and physically meaningful \cite{breuer2016,vega2017,zaccone2025,coppola2025}.}

In the present work, which extends the analysis of Ref. \cite{giordano2025}, we firstly construct a generalized classical Langevin equation \cite{langevin}, in which both friction and noise appear symmetrically in the Hamiltonian formalism.
 By using the Fokker–Planck methodology \cite{fokker,planck}, we derive a generalized Klein–Kramers equation \cite{klein,kramers}, expressed in terms of Poisson brackets, and show that
the asymptotic solution of this equation corresponds to the classical canonical distribution.
Moreover, by introducing the definitions of heat \cite{seki1,seki2}, and entropy \cite{sei1,sei2,sei3}, along a single trajectory, we demonstrate that this approach is fully consistent with the first and second laws of thermodynamics.
It is worth noting that such a formalism, based on the theory of stochastic processes \cite{vankampen,risken,coffey,schuss1,schuss2}, is now systematically applied within statistical mechanics \cite{sch,espo1,espo2,tome0,tome1,tome2,ham1,ham2,boso1,boso2,koide},  including holonomic underdamped and overdamped systems \cite{pan,mura,annphys,giorda}.
Secondly, we introduce the quantum analog of the classical theory. 
We build on our previous studies \cite{palla2020,geom,dupont2024,giordano2024}, by means of which we showed that noise in quantum mechanics always enters multiplicatively, and demonstrated  that canonical quantization -- i.e., replacing Poisson brackets with commutator \cite{landauq,cohen} -- constitutes a rigorous procedure in the classical-to-quantum transition \cite{giordano2025}.
This approach has also been discussed and/or applied in various studies \cite{bianco,dekker1984,sandulescu1987,oliveira2016,oliveira2023,oliveira2024}.
Accordingly, we apply canonical quantization to the Klein–Kramers equation, which is based on the symmetric inclusion of friction and noise in the Hamiltonian equations.
This procedure necessitates the introduction of certain friction operators, which can be defined in two distinct ways depending on whether Hermiticity is imposed or not.
By following both approaches in parallel, we thus obtain two different master equations to describe the nonequilibrium statistical mechanics of a quantum system. 
For both master equations, we analyze their behavior for a simple system, such as the harmonic oscillator.
This analysis, based on Refs. \cite{dekker1984,sandulescu1987}, shows that the resulting equations are of the Lindblad type, i.e. they are completely positive, only if both Hamiltonian equations include friction and noise terms.
This fully justifies the initial choice made in reconstructing the classical statistical mechanics, where friction and noise were introduced in both Hamiltonian equations.
Furthermore, we demonstrate that in the quantum case the two always-positive friction coefficients must also satisfy an additional inequality to ensure complete positivity.
Remarkably, this inequality is identical for both the Hermitian and non-Hermitian operator approaches, suggesting that it could be a more general  feature rather than a consequence of the specific choice of friction operators.
In the following, we show how quantum thermodynamics can be developed starting from the master equation, leading to a novel formulation of the first and second laws.
The first law is based on a generalized definition of heat at the quantum level, while the second law asserts the positivity of entropy production, highlighting its deep connection with the monotonicity of quantum relative entropy \cite{lindblad1974,lindblad1975}.
It should be noted that multiple approaches have been developed to investigate the positivity of entropy production in the quantum context \cite{spohn1978,lebowitz1978,trushechkin2017,trushechkin2019,ruelle2001,pillet2002,kosloff2019}.

Finally, we implemented our formalism to describe the nonequilibrium dynamics of the harmonic oscillator. 
To elucidate our results, we compared five different models.
The first consists of the full equation, with friction and noise included in both Hamiltonian-Langevin equations, using Hermitian friction operators.
The second is the full equation with friction and noise in both Langevin–Hamilton equations, but with non-Hermitian friction operators.
The third and fourth models include friction and noise only in the first Hamiltonian equation, as done classically, with Hermitian and non-Hermitian friction operators, respectively.
Finally, the fifth model corresponds to the Caldeira–Leggett approach \cite{caldeira1981,caldeira1983,caldeira1983bis}, in which friction and noise act only on the first equation, and the mathematical form of the friction operator is further approximated.
The numerical results show that the first two models are fully consistent with thermodynamics and preserve the positivity of the density matrix at all times. This finding is fully in line with the theoretical framework developed in the first part of this work.
The third and fourth models, discussed respectively in Ref. \cite{giordano2025} and Refs. \cite{oliveira2016,oliveira2023,oliveira2024}, are thermodynamically consistent but may yield negative eigenvalues of the density matrix in certain situations. This occurs for specific initial conditions, such as pure states defined by the energy eigenbasis. Consequently, these two models constitute reasonable approximations only for initial mixed states that ensure positivity is preserved throughout the evolution.
Finally, it is straightforward to verify that the Caldeira–Leggett model violates thermodynamic consistency and can also generate negative eigenvalues during the dynamics.

\section{Generalized classical stochastic thermodynamics}

We develop a generalized classical stochastic thermodynamics by considering a system composed of $N$ particles with masses $m_i$, described by coordinates $\vec r_i$, and linear momenta $\vec p_i$ $(i=1,...,N)$.
The total kinetic energy of the system $K_0$ is written as $K_0 = \sum_{i=1}^{N} \frac{1}{2m_i} \vec p_i \cdot \vec p_i$, and the potential energy describing the system is given by $V_0=V_0(\vec r_1,...,\vec r_N)$. 
For this system, we can introduce a set of stochastic generalized Hamilton equations in the following form
\begin{eqnarray}
\label{ham}
 {\dot {\vec p}}_i &=& - \frac{\partial V_0}{\partial \vec r_i}-\beta_p \vec p_i + \sqrt{D_p m_i} \vec n_{p,i}(t) +\vec{f}_i(t) , \\ 
 \label{hambis}
 {\dot {\vec r}}_i &=& \frac{1}{m_i}{\vec p}_i-\beta_q m_i\frac{\partial V_0}{\partial \vec r_i}+\sqrt{D_q m_i} \vec n_{q,i}(t),
\end{eqnarray}
where $\beta_p=0$ and $\beta_q=0$ are two friction coefficients, and  $D_p=0$ and $D_q=0$ are two diffusion constants.
When $\beta_q=0$ and $D_q=0$, we retrieve the classical Langevin model, where each particle is subjected to a force $\vec F_i =- \frac{\partial V_0}{\partial \vec r_i} - \beta_p \vec p_i + \sqrt{D_p m_i} \vec n_{p,i}(t)+\vec{f}_i(t)$, which includes: 
(i) a conservative force field describing the system structure, (ii) an external force field representing the work done on the system (in addition to time, forces $\vec{f}_i $ can also depend on the positions of the particles $\vec r_1,...,\vec r_N$), (iii) a friction force (coefficient $\beta_p$) mimicking the energy transfer from the particles to the thermal bath, and (iv) a noise term (diffusion coefficient $D_p$) mimicking the energy transfer from the bath to the system. 
The friction and noise forces (iii) and (iv) represent the so-called Langevin thermal bath \cite{langevin}. The system of the Hamilton equations with $\beta_q=0$ and $D_q=0$ is consistent with both equilibrium and out-of-equilibrium thermodynamics and statistical mechanics \cite{annphys,giorda}. 
We want to prove here that the additional terms with coefficients $\beta_q\neq 0$ and $D_q\neq 0$ leave these important properties unaltered. This result is particularly relevant in quantum mechanics.
We assume the following hypotheses on the noises: $\vec n_{p,i}(t) \in \mathbb{R}^3$ and $\vec n_{q,i}(t) \in \mathbb{R}^3$ are Gaussian stochastic processes with expectation value $\mathbb{E}\{ \vec n_{p,i}(t)\} = 0$, $\mathbb{E}\{ \vec n_{q,i}(t)\} = 0$ and correlation $\mathbb{E}\{\vec n_{p,i}(t_1) \otimes \vec n_{p,j}(t_2)\} = 2 \delta_{ij}{I}_3 \delta(t_1-t_2)$, $\mathbb{E}\{\vec n_{q,i}(t_1) \otimes \vec n_{q,j}(t_2)\} = 2 \delta_{ij}{I}_3 \delta(t_1-t_2)$, and $\mathbb{E}\{\vec n_{q,i}(t_1) \otimes \vec n_{p,j}(t_2)\} = 0$. 
Here, $\delta_{ij}$ is the Kronecker delta, $\delta(.)$ is the Dirac delta function, $\otimes$ is the tensor product of vectors, and ${I}_3$ is the $3\times3$ identity matrix \cite{vankampen,risken,coffey}. 
If we introduce the Hamiltonian function of the system as $\mathcal{H}_0=K_0 +V_0$, we can rewrite the Hamilton equation in the following more symmetrical form
\begin{eqnarray}
\label{ham1}
 {\dot {\vec p}}_i &=& - \frac{\partial \mathcal{H}_0}{\partial \vec r_i}-m_i\beta_p \frac{\partial \mathcal{H}_0}{\partial \vec p_i}+ \sqrt{D_p m_i} \vec n_{p,i}(t) +\vec{f}_i(t) , \\
 \label{hambis1}
 {\dot {\vec r}}_i &=& \frac{\partial \mathcal{H}_0}{\partial \vec p_i}-m_i\beta_q \frac{\partial \mathcal{H}_0}{\partial \vec r_i}+\sqrt{D_q m_i} \vec n_{q,i}(t).
\end{eqnarray}
The symmetry of this Hamiltonian form justifies the choice made to have a form of friction in Eq. (\ref{hambis}) proportional to $\frac{\partial V_0}{\partial \vec r_i}=\frac{\partial \mathcal{H}_0}{\partial \vec r_i}$.
Further examples of generalized Hamilton–Langevin formulations are available in the existing literature \cite{ham1,ham2,boso1,boso2,koide}.
From a mathematical point of view, our peculiar writing of the Hamilton equations represents a stochastic differential problem with additive noise \cite{schuss1,schuss2}.
We can then apply the Fokker-Planck methodology, which is briefly recalled here for an arbitrary interpretation of the stochastic calculus \cite{vankampen,risken,coffey}. 
Although this distinction is strictly speaking not relevant here since the classical Hamilton equations have an additive noise, we underline that the corresponding quantum equations have multiplicative noise, as thoroughly discussed in Refs. \cite{bianco,giordano2025}. 
We thus consider the stochastic differential system
\begin{eqnarray}
 \label{stoch}
 \frac{dx_i }{dt} = h_i(\vec x,t) + \sum_{j=1}^\mathfrak{m} g_{ij}(\vec x,t) n_j(t),
\end{eqnarray}
with $\mathfrak{n}$ equations and $\mathfrak{m}$ noise terms ($\forall i = 1,..., \mathfrak{n}, \forall j = 1,..., \mathfrak{m}$).
It assumes a precise meaning only after declaring the adopted interpretation of the stochastic calculus. To achieve this, we must specify the parameter $\alpha$, with $0\le \alpha \le 1$, that defines the position of the point at which we calculate any integrated function in the small intervals of the adopted Riemann sum. 
The Gaussian noises $n_j(t)$ ($\forall j = 1,..., \mathfrak{m}$) satisfy the properties $\mathbb{E}\{  n_j(t)\} = 0$, and $\mathbb{E}\{ n_i(t_1)  n_j(t_2)\} = 2 \delta_{ij} \delta(t_1-t_2) $. 
This stochastic differential equation corresponds to the following Fokker-Planck equation for the probability density  $W(\vec x,t)$ \cite{risken,coffey}
\begin{eqnarray}
 \label{fokpla}
 \frac{\partial W(\vec x,t)}{\partial t} &=& -  \sum_{i=1}^\mathfrak{n} \frac{\partial}{\partial x_i} [h_i W(\vec x,t)]-  \sum_{i=1}^\mathfrak{n} \frac{\partial}{\partial x_i} \left\{ 2\alpha\left[ \sum_{k=1}^\mathfrak{n} \sum_{j=1}^\mathfrak{m} g_{kj} \frac{\partial g_{ij}}{\partial x_k}   \right]  W(\vec x,t) \right\} \\ \nonumber
 &&+   \sum_{i=1}^\mathfrak{n} \sum_{j=1}^\mathfrak{m} \frac{\partial^2 }{\partial x_i \partial x_j} \left\{ \left[ \sum_{k=1}^\mathfrak{m} g_{ik} g_{jk}  \right]  W(\vec x,t) \right\},
\end{eqnarray}
where the first term represents the drift, the second is the noise-induced drift (which depends on $\alpha$, and is zero when $\frac{\partial g_{ij}}{\partial x_k}=0$), and the third  
the diffusion (characterizing the effect of the noise). This expression includes the It\^{o} ($\alpha=0$) \cite{itofp}, the Fisk-Stratonovich ($\alpha=1/2$) \cite{fisk,stratofp} and the H\"{a}nggi-Klimontovich ($\alpha=1$) \cite{hanggifp,klimofp} interpretations, as particular cases (see Ref.\cite{sokolov}).

Eventually, we can write the Fokker-Planck \cite{fokker,planck} (or Klein-Kramers \cite{klein,kramers}) equation associated with Eqs.(\ref{ham}) in the following form 
\begin{eqnarray}
\nonumber
 \frac{\partial W}{\partial t} &=&  -\sum_{i=1}^N \frac{{\vec p}_i}{m_i}\cdot \frac{\partial W}{\partial {\vec r}_i} +\sum_{i=1}^N  \frac{\partial V}{\partial {\vec r}_i} \cdot \frac{\partial W}{\partial {\vec p}_i} -\sum_{i=1}^N   {\vec f}_i \cdot \frac{\partial W}{\partial {\vec p}_i}\\
 \nonumber
 &&+3N \beta_p W + \beta_p \sum_{i=1}^N{\vec p}_i\cdot   \frac{\partial W}{\partial{\vec p}_i}+ D_p\sum_{i=1}^N m_i \frac{\partial^2 W }{\partial {\vec p}_i ^2} \\  \label{klein1}
 && +\beta_q\sum_{i=1}^N m_i\frac{\partial^2 V_0 }{\partial {\vec r}_i ^2}W+\beta_q\sum_{i=1}^N m_i\frac{\partial V_0}{\partial {\vec r}_i} \cdot \frac{\partial W}{\partial {\vec r}_i}+ D_q\sum_{i=1}^N m_i \frac{\partial^2 W }{\partial {\vec r}_i ^2},
\end{eqnarray}
where $W=W({\vec r}_1,...,{\vec r}_N,{\vec p}_1,...,{\vec p}_N,t)$. The derivative $\frac{\partial^2 W }{\partial {\vec p}_i ^2}$ (or $\frac{\partial^2 W }{\partial {\vec r}_i ^2}$) represents the Laplacian operator with respect to the three components of ${\vec p}_i$ (or ${\vec r}_i$).

A more interesting form for the following developments can be found by introducing the Poisson brackets as follows \cite{gantmacher}
\begin{eqnarray}
\nonumber
 &&\frac{\partial W}{\partial t} = \left\lbrace \mathcal{H},W \right\rbrace  -\sum_{i=1}^N   {\vec f}_i \cdot \frac{\partial W}{\partial {\vec p}_i}\\
 \nonumber
 &&+ \beta_p \sum_{i=1}^N m_i\left( \left\lbrace r_{xi},\frac{\partial \mathcal{H}_0}{\partial  p_{xi}} W\right\rbrace+\left\lbrace r_{yi},\frac{\partial \mathcal{H}_0}{\partial  p_{yi}} W\right\rbrace+\left\lbrace r_{zi},\frac{\partial \mathcal{H}_0}{\partial  p_{zi}} W\right\rbrace \right)\\
 \nonumber
 &&+ D_p \sum_{i=1}^N m_i \left(\left\lbrace r_{xi},  \left\lbrace x_i,W\right\rbrace \right\rbrace+\left\lbrace r_{yi},  \left\lbrace y_i,W\right\rbrace \right\rbrace+\left\lbrace r_{zi},  \left\lbrace z_i,W\right\rbrace \right\rbrace\right)
 \\
 \nonumber
 &&- \beta_q \sum_{i=1}^N m_i\left(\left\lbrace p_{xi},\frac{\partial \mathcal{H}_0}{\partial  r_{xi}} W\right\rbrace+\left\lbrace p_{yi},\frac{\partial \mathcal{H}_0}{\partial  r_{yi}} W\right\rbrace+\left\lbrace p_{zi},\frac{\partial \mathcal{H}_0}{\partial  r_{zi}} W\right\rbrace  \right)\\
 &&+ D_q \sum_{i=1}^N m_i \left(\left\lbrace p_{xi},  \left\lbrace p_{xi},W\right\rbrace \right\rbrace+\left\lbrace p_{yi},  \left\lbrace p_{yi},W\right\rbrace \right\rbrace+\left\lbrace p_{zi},  \left\lbrace p_{zi},W\right\rbrace \right\rbrace\right),
 \label{poisson}
\end{eqnarray}
where $\vec{r}_i=(r_{xi},r_{yi},r_{zi})$, $\vec{p}_i=(p_{xi},p_{yi},p_{zi})$, and $\mathcal{H}_0=K_0 +V_0$. This form is particularly suitable for the application of canonical quantization.

The asymptotic behavior of Eq. (\ref{poisson}) for large times ($t\gg 1/\beta_p$, $t\gg 1/\beta_q$) is characterized by the canonical or Gibbs distribution \cite{gibbs}. Indeed, if the forces $\vec{f}_i$ are absent, and the integral defining the classical partition function,
\begin{eqnarray}
 \label{trenta}
 Z_{cl} = \int_{\mathbb{R}^{3N}} \int_{\mathbb{R}^{3N}} e^{- \frac{1}{k_B T}\mathcal{H}_0(\vec{q},\vec{p})} d\vec{q} d\vec{p},
\end{eqnarray}
is convergent (with 
$\vec{q}=(\vec{r}_1,...,\vec{r}_N) \in \mathbb{R}^{3N}$  and $\vec{p}=(\vec{p}_1,...,\vec{p}_N) \in \mathbb{R}^{3N}$), then the asymptotic solution of Eq. (\ref{poisson}) is given by the Gibbs distribution in 
phase space
\begin{eqnarray}
 \label{trentuno}
 W_{eq}(\vec{q},\vec{p}) = \frac{1}{Z_{cl}} e^{- \frac{1}{k_B T}\mathcal{H}_0(\vec{q},\vec{p})},
\end{eqnarray}
as can be easily proved by substitution. 
This asymptotic solution allows identification of the diffusion constants through the expressions $D_p =  k_B T \beta_p$ and $D_q =  k_B T \beta_q$, referred to as the classical Einstein fluctuation-dissipation relations \cite{coffey,risken}. 
This means that our formalism is consistent with equilibrium thermodynamics, but it must also be verified that it behaves correctly during relaxation. 
To this end, in the following we show how the two laws of thermodynamics can be rederived out of equilibrium.

We define the internal energy $\mathcal{E}$ of the system as the average value, with respect of the probability density defined by Eq.(\ref{klein1}) or (\ref{poisson}), of the sum of kinetic energy and potential energy $\mathcal{E}=\mathbb{E} \left\lbrace K_0+V_0\right\rbrace$, and we calculate the rate $\frac{d\mathcal{E}}{dt}$. 
After straightforward calculations, we obtain
\begin{eqnarray}
\nonumber
\frac{d\mathcal{E}}{dt}&=&\sum_{i=1}^N\vec{f}_i\cdot\frac{\mathbb{E}\left\lbrace \vec{p}_i\right\rbrace}{m_i}+\beta_p\left(k_B T\sum_{i=1}^Nm_i\mathbb{E}\left\lbrace \frac{\partial^2 \mathcal{H}_0 }{\partial {\vec p}_i ^2} \right\rbrace - \sum_{i=1}^Nm_i\mathbb{E}\left\lbrace \frac{\partial \mathcal{H}_0}{\partial {\vec p}_i } \cdot\frac{\partial \mathcal{H}_0 }{\partial {\vec p}_i }\right\rbrace\right)\\
\nonumber
&&+\beta_q\left(k_B T\sum_{i=1}^Nm_i\mathbb{E}\left\lbrace \frac{\partial^2 \mathcal{H}_0 }{\partial {\vec r}_i ^2} \right\rbrace - \sum_{i=1}^Nm_i\mathbb{E}\left\lbrace \frac{\partial \mathcal{H}_0 }{\partial {\vec r}_i } \cdot\frac{\partial \mathcal{H}_0 }{\partial {\vec r}_i }\right\rbrace \right)\\
\nonumber
&=&\sum_{i=1}^N\vec{f}_i\cdot\frac{\mathbb{E}\left\lbrace \vec{p}_i\right\rbrace}{m_i}+2\beta_p\left(\frac{3}{2}
N k_B T -\mathbb{E}\left\lbrace K_0\right\rbrace\right)\\
\nonumber
&&+\beta_q\left(k_B T\sum_{i=1}^Nm_i\mathbb{E}\left\lbrace \frac{\partial^2 V_0 }{\partial {\vec r}_i ^2} \right\rbrace - \sum_{i=1}^Nm_i\mathbb{E}\left\lbrace \frac{\partial V_0 }{\partial {\vec r}_i } \cdot\frac{\partial V_0 }{\partial {\vec r}_i }\right\rbrace \right)\\
&=&\frac{d\mathbb{E}\left\lbrace L \right\rbrace}{dt}+
\frac{d\mathbb{E}\left\lbrace Q \right\rbrace}{dt}.
\label{1pr}
\end{eqnarray}
This expression represents the first law of thermodynamics,
from which we can identify the rate of average work $\frac{d\mathbb{E}\left\lbrace L \right\rbrace}{dt}$ done on the system with the average power $\sum_{i=1}^N\vec{f}_i\cdot\mathbb{E}\left\lbrace \vec{p}_i\right\rbrace/m_i$, and the remaining two terms, proportional to $\beta_p$ and $\beta_q$ respectively, with the rate of average heat $\frac{d\mathbb{E}\left\lbrace Q \right\rbrace}{dt}$ entering the system.

Then, we prove that the heat flux is zero when the following generalized equipartition of energy is satisfied. At equilibrium, we can write
\begin{eqnarray}
\nonumber
    \mathbb{E}\left\lbrace \frac{\partial^2 \mathcal{H}_0 }{\partial {\vec s}_i ^2} \right\rbrace &=&\frac{1}{Z_{cl} }\int_{\mathbb{R}^{3N}} \int_{\mathbb{R}^{3N}} \frac{\partial^2 \mathcal{H}_0 }{\partial {\vec s}_i ^2}  e^{- \frac{1}{k_B T}\mathcal{H}_0(\vec{q},\vec{p})} d\vec{q} d\vec{p}\\
    \nonumber
    &=&\frac{1}{k_B T }\frac{1}{Z_{cl} }\int_{\mathbb{R}^{3N}} \int_{\mathbb{R}^{3N}} \frac{\partial \mathcal{H}_0 }{\partial {\vec s}_i } \cdot\frac{\partial \mathcal{H}_0 }{\partial {\vec s}_i }  e^{- \frac{1}{k_B T}\mathcal{H}_0(\vec{q},\vec{p})} d\vec{q} d\vec{p}\\
    &=&\frac{1}{k_B T }\mathbb{E}\left\lbrace \frac{\partial \mathcal{H}_0}{\partial {\vec s}_i } \cdot\frac{\partial \mathcal{H}_0 }{\partial {\vec s}_i }\right\rbrace,
    \label{equic}
\end{eqnarray}
where $s=p$ or $r$, and a multidimensional integration by parts was used. This shows that both terms in the definition of heat flow tend toward zero when thermodynamic equilibrium is reached. 
Indeed, for $s=p$ the classical equipartition theorem is obtained in the form $\mathbb{E}\left\lbrace K_0\right\rbrace=\frac{3}{2}
N k_B T$, and when $s=r$, we get the new dual expression $ \sum_{i=1}^Nm_i\mathbb{E}\left\lbrace \frac{\partial V_0 }{\partial {\vec q}_i } \cdot\frac{\partial V_0 }{\partial {\vec q}_i }\right\rbrace=k_B T\sum_{i=1}^Nm_i\mathbb{E}\left\lbrace \frac{\partial^2 V_0 }{\partial {\vec q}_i ^2} \right\rbrace$. 
It can also be remarked that the heat rate here defined is consistent with the development of stochastic energetics \cite{seki1, seki2}.

To substantiate the previous explicit expressions of the heat rate, we reobtain the second law of thermodynamics by introducing the Gibbs entropy of the system as:
\begin{eqnarray}
\mathcal{S}=-k_B \mathbb{E}\left\lbrace \log W \right\rbrace =-k_B\int_{\mathbb{R}^{3N}}\int_{\mathbb{R}^{3N}}W\log W d\vec{q} d\vec{p}.
\label{entr}
\end{eqnarray}
This expression means that the microscopic (non-averaged) entropy along a given system trajectory is defined as $-k_B  \log W$, consistently with Refs.\cite{sei1,sei2,sei3}.
The evolution equation for $W$ can be rewritten as
\begin{eqnarray}
 \frac{\partial W}{\partial t} = \left\lbrace \mathcal{H}_0,W \right\rbrace -\sum_{i=1}^N   {\vec f}_i \cdot \frac{\partial W}{\partial {\vec p}_i} -\sum_{i=1}^N\frac{\partial \vec{J}_{pi}}{\partial \vec{p}_i} -\sum_{i=1}^N\frac{\partial \vec{J}_{ri}}{\partial \vec{r}_i},
 \label{fpf}
\end{eqnarray}
where we introduced:
\begin{eqnarray}
\label{fluxp}
\vec{J}_{pi}&=&-\beta_p W \vec{p}_i-k_BT\beta_p m_i\frac{\partial W}{\partial \vec{p}_i}=-\beta_p m_i W \frac{\partial \mathcal{H}_0}{\partial \vec{p}_i}-k_BT\beta_p m_i\frac{\partial W}{\partial \vec{p}_i},\\
\label{fluxq}
\vec{J}_{ri}&=&-\beta_q m_i W \frac{\partial V_0}{\partial \vec{r}_i}-k_BT\beta_q m_i\frac{\partial W}{\partial \vec{r}_i}=-\beta_q m_i W \frac{\partial \mathcal{H}_0}{\partial \vec{r}_i}-k_BT\beta_q m_i\frac{\partial W}{\partial \vec{r}_i},
\end{eqnarray}
which are the fluxes in momentum space and in configurational space, respectively. 
We note that the derivatives $\frac{\partial \vec{J}_{pi}}{\partial \vec{p}_i}$ and $\frac{\partial \vec{J}_{ri}}{\partial \vec{r}_i}$ represent the divergence with respect to the components of $\vec{p}_i$ and $\vec{r}_i$, respectively, giving the Fokker-Planck equation the form of a continuity equation.
The total entropy rate can be written in the form:
\begin{eqnarray}
\frac{d\mathcal{S}}{dt}=-k_B\int_{\mathbb{R}^{3N}}\int_{\mathbb{R}^{3N}}\frac{\partial W}{\partial t}\log W d\vec{q} d\vec{p},
\label{entrate}
\end{eqnarray}
in which Eq.(\ref{fpf}) can be substituted for $\partial W/\partial t$. It can then be verified that the Liouville term and the term with external forces (depending solely on positions and time) have zero entropic contribution. Therefore: 
\begin{eqnarray}
\nonumber
\frac{d\mathcal{S}}{dt}&=&k_B\int_{\mathbb{R}^{3N}}\int_{\mathbb{R}^{3N}} \left(\sum_{i=1}^N\frac{\partial \vec{J}_{pi}}{\partial \vec{p}_i} +\sum_{i=1}^N\frac{\partial \vec{J}_{ri}}{\partial \vec{r}_i}\right)\log W d\vec{q} d\vec{p}\\
&=&-k_B\int_{\mathbb{R}^{3N}}\int_{\mathbb{R}^{3N}}\frac{1}{W} \left(\sum_{i=1}^N\vec{J}_{pi}\cdot\frac{\partial W}{\partial \vec{p}_i} +\vec{J}_{ri}\cdot\sum_{i=1}^N\frac{\partial W }{\partial \vec{r}_i}\right) d\vec{q} d\vec{p}.
\label{entrateb}
\end{eqnarray}
We can now express the derivatives $\frac{\partial W}{\partial \vec{p}_i}$ and $\frac{\partial W}{\partial \vec{r}_i}$ as a function of the fluxes $\vec{J}_{pi}$ and $\vec{J}_{ri}$, using Eqs. (\ref{fluxp}) and (\ref{fluxq}).
This procedure leads to the entropy balance in the form:
\begin{eqnarray}
\nonumber
\frac{d\mathcal{S}}{dt}&=&\frac{1}{T} \frac{d\mathbb{E}\left\lbrace Q\right\rbrace }{dt}+\frac{1}{\beta_p T} \int_{\mathbb{R}^{3N}}\int_{\mathbb{R}^{3N}}\frac{1}{W}\sum_{i=1}^N\frac{\vec{J}_{pi}\cdot\vec{J}_{pi}}{m_i } d\vec{q} d\vec{p}\\
&&+\frac{1}{\beta_q T} \int_{\mathbb{R}^{3N}}\int_{\mathbb{R}^{3N}}\frac{1}{W}\sum_{i=1}^N\frac{\vec{J}_{ri}\cdot\vec{J}_{ri}}{m_i } d\vec{q} d\vec{p},
\label{2pr}
\end{eqnarray}
where
\begin{eqnarray}
     \frac{d\mathbb{E}\left\lbrace Q\right\rbrace }{dt}&=&\beta_p\left(k_B T\sum_{i=1}^Nm_i\mathbb{E}\left\lbrace \frac{\partial^2 \mathcal{H}_0 }{\partial {\vec p}_i ^2} \right\rbrace - \sum_{i=1}^Nm_i\mathbb{E}\left\lbrace \frac{\partial \mathcal{H}_0}{\partial {\vec p}_i } \cdot\frac{\partial \mathcal{H}_0 }{\partial {\vec p}_i }\right\rbrace\right)\\
\nonumber
&&+\beta_q\left(k_B T\sum_{i=1}^Nm_i\mathbb{E}\left\lbrace \frac{\partial^2 \mathcal{H}_0 }{\partial {\vec r}_i ^2} \right\rbrace - \sum_{i=1}^Nm_i\mathbb{E}\left\lbrace \frac{\partial \mathcal{H}_0 }{\partial {\vec r}_i } \cdot\frac{\partial \mathcal{H}_0 }{\partial {\vec r}_i }\right\rbrace \right),
\end{eqnarray}
consistent with the analysis carried out above, to rederive the first law of thermodynamics.
In Eq. (\ref{2pr}), we identify the first term with the entropy flow (heat-driven disorder), and the sum of the second and third terms with the entropy production (irreversibility). 
It may be noted that the entropy production is always non-negative since it is constituted by positive-definite quadratic expressions \cite{sch,espo1,espo2,tome0,tome1,tome2,giorda}.
Therefore,  the second law of thermodynamics is reobtained in the classical form:  
\begin{eqnarray}
\frac{d\mathcal{S}}{dt}& \geq &\frac{1}{T} \frac{d\mathbb{E}\left\lbrace Q\right\rbrace }{dt},
\label{2prfin}
\end{eqnarray}
where the equality is satisfied only for quasi-static transformations, evolving not far from the thermodynamic equilibrium \cite{giorda}. 
It is interesting to note that this thermodynamic structure is preserved even when we treat a holonomic system with arbitrary mechanical constraints \cite{annphys,giorda}. 
Moreover, this scheme can be further generalized to introduce the overdamped approximation \cite{annphys,giorda,pan}, and multiple reservoirs \cite{mura}.

\section{Density matrix evolution in quantum mechanics}

 Firstly, the concept of density matrix in quantum mechanics is briefly recalled. A mixed state for an arbitrary system is described by $M$ wave-functions $\Psi_1,...,\Psi_M$, associated with the corresponding probabilities $p_1,...,p_M$, with $\sum_{j=1}^Mp_j=1$.
The density operator is thereby defined as $\varrho(\vec{q},\vec{q}',t)=\sum_{j=1}^Mp_j\Psi_j(\vec{q},t)\Psi_j^*(\vec{q}',t)$. 
The 
expectation value of an observable  $f$ is then calculated as $\mathbb{E}\left\lbrace f\right\rbrace=\sum_{j=1}^Mp_j\langle \Psi_j\vert f \Psi_j \rangle=\int (f\varrho)\vert_{\vec{q}'=\vec{q}}d\vec{q}$, where $f$ represents the quantum operator acting only on the variables $\vec{q}$.
By adopting the orthonormal basis $\left\lbrace \varphi_n(\vec{q}):\mathbb{R}^{3N}\to\mathbb{C} \right\rbrace$, we have that $\Psi_j=a_{kj}\varphi_k$, with $a_{kj}=\langle \varphi_k\vert  \Psi_j \rangle$ (Einstein's summation convention is assumed).
Hence, the expectation value of $f$ can be written as $\mathbb{E}\left\lbrace f\right\rbrace=\int \sum_{j=1}^Mp_ja_{kj}f\varphi_ka_{hj}^*\varphi_h^*d\vec{q}=\varrho_{kh}f_{hk}=\mbox{Tr}(\varrho f)$, where we identified the density matrix representation $\varrho_{kh}=\sum_{j=1}^Mp_ja_{kj}a_{hj}^*$, and the operator representation $f_{hk}=\int \varphi_h^*f\varphi_kd\vec{q}=\langle \varphi_h\vert f \varphi_k \rangle$.

The density matrix $\varrho_{kh}$ satisfies certain properties that will have to be fulfilled also during the time evolution \cite{landauq,cohen}: 
(i)  its trace is unitary,  $\mbox{Tr}\varrho=\varrho_{kk}=\sum_{j=1}^Mp_ja_{kj}a_{kj}^*=\sum_{j=1}^Mp_j\langle \Psi_j\vert  \Psi_j \rangle=1$; 
(ii) the diagonal elements 
are non-negative, $\varrho_{kk}=\sum_{j=1}^Mp_ja_{kj}a_{kj}^*=\sum_{j=1}^Mp_j\vert\langle \varphi_k\vert  \Psi_j \rangle\vert^2\ge 0$ (without the sum over $k$); 
(iii) the density matrix $\varrho_{kh}=\sum_{j=1}^Mp_ja_{kj}a_{hj}^*$ is Hermitian,  $\varrho_{hk}^*=\sum_{j=1}^Mp_ja_{hj}^*a_{kj}=\varrho_{kh}$, or $\varrho^{T*}=\varrho$ ($T$ means ``transposed''); 
(iv) the density matrix is positive-definite,  $v^{T*}\varrho v=v_k^*\varrho_{kh}v_h=\sum_{j=1}^Mp_jv_k^*a_{kj}v_ha_{hj}^*=\sum_{j=1}^Mp_j(v_ka_{kj}^*)^*(v_ha_{hj}^*)=\sum_{j=1}^Mp_j\vert v_ka_{kj}^*\vert^2>0$.

{  Here, for the sake of clarity and simplicity, all properties are presented assuming a finite ensemble of $M$ pure states. We emphasize, however, that the formalism is fully general and can be extended to more general statistical mixtures. In particular, the derivations remain valid for countably infinite ensembles, where the sum over the states extends to infinity, provided the probabilities converge so that the density matrix remains trace-class. Furthermore, the results also hold for continuous ensembles, in which the discrete sum is replaced by an integral over a continuous set of states. In both cases, the key properties of the density matrix — Hermiticity, positivity, and unit trace — are preserved, ensuring that all conclusions drawn for finite ensembles carry over straightforwardly to these more general settings.}

For a conservative system defined by the Hamiltonian $\mathcal{H}_0$, the time evolution of the density matrix is described by the Liouville-von Neumann equation  \cite{landauq,cohen}
\begin{equation}
    \frac{d\varrho}{dt}=\frac{1}{i\hbar}\left[\mathcal{H}_0,\varrho\right],
    \label{neumann}
\end{equation}
where $\left[\mathcal{A},\mathcal{B}\right]=\mathcal{A}\mathcal{B}-\mathcal{B}\mathcal{A}$ is the commutator of operators $\mathcal{A}$ and $\mathcal{B}$.
In the following Sections, we will discuss the modifications made to this conservative equation in order to take into account the interaction with a thermal bath. 
This procedure can be conducted by using Hermitian friction operators (as discussed in Section \ref{herm}) or non-Hermitian friction operators (see Section \ref{nonherm}). In both cases, a master equation for open quantum systems will be obtained.

\section{First approach: Hermitian friction operators}
\label{herm}

To introduce noise and dissipation terms into this equation, we exploit the analogy with the classical approach.
In particular, we apply the canonical quantization, which replaces Poisson brackets $\left\lbrace \mathcal{A},\mathcal{B}\right\rbrace$ with commutators $\frac{1}{i\hbar}\left[\mathcal{A},\mathcal{B}\right]$ \cite{landauq,cohen}.
Following this principle, we develop the quantum counterpart of Eq. (\ref{poisson}).

The terms on the third and fifth lines, containing double Poisson brackets, are easily converted to the quantum case by using a double commutator. The quantization of these terms is rigorous in that it corresponds to the average with respect to a stochastic quantum Hamiltonian, as exhaustively described in Refs. \cite{bianco,giordano2025}.
 This development involves a multi-dimensional geometric Brownian motion, see Appendix A in Ref. \cite{giordano2025}, which is a generalization of the simpler case discussed in Ref. \cite{geom}.

Regarding the second line,  the terms $m_i\frac{\partial \mathcal{H}_0}{\partial  p_{xi}} W$, $m_i\frac{\partial \mathcal{H}_0}{\partial  p_{yi}} W$, and $m_i\frac{\partial \mathcal{H}_0}{\partial  p_{zi}} W$ must be transformed into Hermitian operators. 
Similarly, in the fourth line the terms $m_i\frac{\partial \mathcal{H}_0}{\partial r_{xi}} W$, $m_i\frac{\partial \mathcal{H}_0}{\partial r_{yi}} W$, and $m_i\frac{\partial \mathcal{H}_0}{\partial r_{zi}} W$ must also be transformed into Hermitian operators. 

Since the product of two Hermitian operators is not necessarily Hermitian, but their symmetrization is always Hermitian,
we substitute the terms in the second line with the quantum symmetrizations  $\frac{1}{2}(\Theta^p_{xi}{\varrho}+{\varrho}\Theta^p_{xi})$, $\frac{1}{2}(\Theta^p_{yi}{\varrho}+{\varrho}\Theta^p_{yi})$, and $\frac{1}{2}(\Theta^p_{zi}{\varrho}+{\varrho}\Theta^p_{zi})$, where $\Theta^p_{xi}$, $\Theta^p_{yi}$, and $\Theta^p_{zi}$ are Hermitian operators to be determined ($\forall i=1..N$), taking the role of classical momenta $p_{xi},p_{yi}$, and $p_{zi}$.
Similarly, we substitute the terms in the fourth line with the quantum symmetrizations  $\frac{1}{2}(\Theta^q_{xi}{\varrho}+{\varrho}\Theta^q_{xi})$, $\frac{1}{2}(\Theta^q_{yi}{\varrho}+{\varrho}\Theta^q_{yi})$, and $\frac{1}{2}(\Theta^q_{zi}{\varrho}+{\varrho}\Theta^q_{zi})$, where $\Theta^q_{xi}$, $\Theta^q_{yi}$, and $\Theta^q_{zi}$ are Hermitian operators to be determined ($\forall i=1..N$), taking the role of classical terms $m_i\frac{\partial V_0}{\partial r_{xi}}$, $m_i\frac{\partial V_0}{\partial r_{yi}}$, and $m_i\frac{\partial V_0}{\partial r_{zi}}$.

By considering all these terms, added to Eq. (\ref{neumann}), we obtain the complete equation
\noindent
\begin{equation}
\label{maindiffforces}
  \makebox[0pt][l]{\hspace*{-\mathindent}\(
    \begin{array}{rcl}
\displaystyle   \frac{d\varrho}{dt}&=&\displaystyle\frac{1}{i\hbar}\left[\mathcal{H}_0,\varrho\right]-\frac{1}{i\hbar}\sum_{k=1}^{N}\sum_{s=x,y,z}f_{sk}\left[r_{sk},\varrho\right]\\
    &-&\displaystyle\frac{k_B T \beta_p}{\hbar^2}\sum_{k=1}^{N}m_k\left(\sum_{s=x,y,z}\left[r_{sk},\left[r_{sk},\varrho\right]\right]\right)+\frac{\beta_p}{2i\hbar}\sum_{k=1}^{N}\left(\sum_{s=x,y,z}\left[r_{sk},\Theta^p_{sk}\varrho+\varrho\Theta^p_{sk}\right]\right)\\
    &-&\displaystyle\frac{k_B T \beta_q}{\hbar^2}\sum_{k=1}^{N}m_k\left(\sum_{s=x,y,z}\left[p_{sk},\left[p_{sk},\varrho\right]\right]\right)-\frac{\beta_q}{2i\hbar}\sum_{k=1}^{N}\left(\sum_{s=x,y,z}\left[p_{sk},\Theta^q_{sk}\varrho+\varrho\Theta^q_{sk}\right]\right),
    \end{array}
  \)}\hfill
\label{mastereq1}
\end{equation}
which represents the master equation for the evolution of the density matrix.

Now we must find the mathematical form of the friction Hermitian operators $\Theta^{p,q}_{sk}$, $s=x,y,z$, $k=1,...,N$, in such a way that the asymptotic behavior of the equation, without external forces $f_{sk}$, is described by the canonical quantum distribution
\begin{equation}
    \lim_{t\to\infty}{\varrho}=\varrho_{eq}=\frac{1}{Z_{qu}}e^{-\frac{\mathcal{H}_0}{k_BT}},
    \label{boltz}
\end{equation}
where $Z_{qu}$ is the quantum partition function
\begin{equation}
 Z_{qu}=\mbox{Tr}\left(e^{-\frac{\mathcal{H}_0}{k_BT}}\right). 
\end{equation}
We impose this asymptotic quantum canonical distribution on Eq. (\ref{maindiffforces}) (with $f_{sk}=0$), obtaining the relations:
\begin{eqnarray}
    \frac{2im_k k_B T}{\hbar}\left[r_{sk},e^{-\frac{\mathcal{H}_0}{k_BT}}\right]&=&\Theta^p_{sk}e^{-\frac{\mathcal{H}_0}{k_BT}}+e^{-\frac{\mathcal{H}_0}{k_BT}}\Theta^p_{sk},
    \label{mainequationp}
\\
   -\frac{2im_k k_B T}{\hbar}\left[p_{sk},e^{-\frac{\mathcal{H}_0}{k_BT}}\right]&=&\Theta^q_{sk}e^{-\frac{\mathcal{H}_0}{k_BT}}+e^{-\frac{\mathcal{H}_0}{k_BT}}\Theta^q_{sk},
    \label{mainequationq}
\end{eqnarray}
for $s=x,y,z$, and $k=1,...,N$.
From a mathematical point of view, these equations in $\Theta^{p,q}_{sk}$ are matrix equations of the form ${A}{X}+{X}{A}={C}$, which are sometimes called the Sylvester or Lyapunov equations, see e.g. Refs.\cite{gant,lanc}. 
We proved in Ref. \cite{giordano2025} that this equation has a unique solution
\begin{eqnarray}
{X}=-\int_0^{+\infty}e^{{A}\xi}{C}e^{{A}\xi}d\xi,
\end{eqnarray}
provided that ${A}$ has all eigenvalues with negative real part. 
Hence, we can write the explicit solutions
of Eqs. (\ref{mainequationp}) and (\ref{mainequationq}) as:
\begin{eqnarray}
    \Theta^p_{sk}&=&\frac{2im_k k_B T}{\hbar}\int_0^{+\infty}
e^{-\xi e^{-\frac{\mathcal{H}_0}{k_BT}}}\left[r_{sk},e^{-\frac{\mathcal{H}_0}{k_BT}}\right]e^{-\xi e^{-\frac{\mathcal{H}_0}{k_BT}}}d\xi,
\label{integp}\\
\Theta^q_{sk}&=&-\frac{2im_k k_B T}{\hbar}\int_0^{+\infty}
e^{-\xi e^{-\frac{\mathcal{H}_0}{k_BT}}}\left[p_{sk},e^{-\frac{\mathcal{H}_0}{k_BT}}\right]e^{-\xi e^{-\frac{\mathcal{H}_0}{k_BT}}}d\xi.
\label{integq}
\end{eqnarray}

These expressions are rather complicated, but they lead to simpler results when projected onto the energy basis of the Hamiltonian operator.
By assuming that the spectrum of the system is discrete and non-degenerate, we can write $\mathcal{H}_0\varphi_n(\vec{q})=E_n\varphi_n(\vec{q})$, with $\langle \varphi_n\vert\varphi_m \rangle=\delta_{nm} $.
In this basis, the operator $e^{-\frac{\mathcal{H}_0}{k_BT}}$ is diagonal with elements $e^{-\frac{E_n}{k_BT}}$. 
To simplify the notation, we introduce the quantity $e_n=e^{-\frac{E_n}{k_BT}}>0$. 
Therefore, the  matrix $[r_{sk},e^{-\frac{\mathcal{H}_0}{k_BT}}]$ is composed of the  elements
$[r_{sk},e^{-\frac{\mathcal{H}_0}{k_BT}}]_{pq}=r_{sk,pq}(e_q-e_p)$, and the  matrix $[p_{sk},e^{-\frac{\mathcal{H}_0}{k_BT}}]$ is composed of the  elements
$[p_{sk},e^{-\frac{\mathcal{H}_0}{k_BT}}]_{pq}=p_{sk,pq}(e_q-e_p)$.

Hence, the structure of the friction operators in Eqs. (\ref{integp}) and (\ref{integq}) takes the manifestly Hermitian form: 
\begin{eqnarray}
\label{frip}
    \Theta^p_{sk,\ell j}&=&\frac{2im_k k_B T}{\hbar}r_{sk,\ell j}\frac{e_j-e_{\ell}}{e_j+e_{\ell}},\\
    \label{friq}
    \Theta^q_{sk,\ell j}&=&-\frac{2im_k k_B T}{\hbar}p_{sk,\ell j}\frac{e_j-e_{\ell}}{e_j+e_{\ell}}.
\end{eqnarray}

To further simplify these results, we use a simple relation between the coefficients $r_{sk,\ell j}=\langle\varphi_{\ell}\vert r_{sk}\varphi_j \rangle$ and $p_{sk,\ell j}=-i\hbar\langle\varphi_{\ell}\vert \frac{\partial}{\partial r_{sk}}\varphi_j \rangle$. 
We start by observing that $[r_{sk},p_{sk}]=i\hbar$, and therefore $[r_{sk},p_{sk}^2]=[r_{sk},p_{sk}]p_{sk}+p_{sk}[r_{sk},p_{sk}]=2i\hbar p_{sk}$. Then, we can write $[r_{sk},\mathcal{H}_0]=[r_{sk},K_0+V_0]=\frac{1}{2m_k}[r_{sk},p_{sk}^2]=\frac{i\hbar}{m_k}p_{sk}$. 
Projecting the latter relationship onto the energy basis of the system, we obtain  $r_{sk,\ell j}=\frac{i\hbar}{m_k}\frac{p_{sk,\ell j}}{E_j-E_{\ell}}$. 
On the other hand, we can also consider the relation  $[p_{sk},\mathcal{H}_0]=[p_{sk},K_0+V_0]=-i\hbar\frac{\partial V_0}{\partial r_{sk}}$. The projection of this equality onto the energy basis leads to $p_{sk,\ell j}=-\frac{i\hbar}{E_j-E_{\ell}}\left(\frac{\partial V_0}{\partial r_{sk}}\right)_{\ell j}$.
Finally, substituting these results and using the property $\frac{e_j-e_{\ell}}{e_j+e_
{\ell}}=\tanh\left(\frac{E_{\ell}-E_j}{2k_B T}\right)$ into Eqs. (\ref{frip}) and (\ref{friq}), we obtain:
\begin{eqnarray}
\label{raprp}
    \Theta^p_{sk,\ell j}&=&p_{sk,\ell j}\frac{\tanh\left(\frac{E_{\ell}-E_j}{2k_B T}\right)}{\frac{E_{\ell}-E_j}{2k_B T}},\\
    \label{raprq}
    \Theta^q_{sk,\ell j}&=&m_k\left(\frac{\partial V_0}{\partial r_{sk}}\right)_{\ell j}\frac{\tanh\left(\frac{E_{\ell}-E_j}{2k_B T}\right)}{\frac{E_{\ell}-E_j}{2k_B T}}.
\end{eqnarray}
These expressions show that the friction operators $\Theta^p_{sk}$ can not be exactly identified with $p_{sk}$, and the friction operators $\Theta^q_{sk}$ are not exactly identified with $m_k\frac{\partial V_0}{\partial r_{sk}}$. 
However, these operators are formally similar to moments and masses multiplied by potential gradients, while they also depend on the energy levels of the system and the temperature itself. 
A generalization to degenerate energy spectra is given in Appendix D of Ref. \cite{giordano2025}.

Another general form of the friction operators was obtained in Ref. \cite{giordano2025}. In that work we  only provided proof for $\Theta^p_{sk}$, but the result can easily be generalized for $\Theta^q_{sk}$, as follows:
\begin{eqnarray}
    \Theta^p_{sk}&=&\frac{2 }{\pi}\int_{-\infty}^{+\infty}
e^{+i\frac{\mathcal{H}_0}{k_BT}\eta}p_{sk}e^{-i\frac{\mathcal{H}_0}{k_BT}\eta}\log\left[\coth\left(\frac{\pi}{2}\vert\eta\vert\right)\right]d\eta,
\label{integ2p}\\
\Theta^q_{sk}&=&\frac{2 }{\pi}\int_{-\infty}^{+\infty}
e^{+i\frac{\mathcal{H}_0}{k_BT}\eta}m_k\frac{\partial V_0}{\partial r_{sk}}e^{-i\frac{\mathcal{H}_0}{k_BT}\eta}\log\left[\coth\left(\frac{\pi}{2}\vert\eta\vert\right)\right]d\eta.
\label{integ2q}
\end{eqnarray}
Of course, the interpretation of these results is similar to the previous one.  
Moreover, these expressions allow friction operators to be written as power series with coefficients $\frac{1}{(k_BT)^{2n}}$, as
\noindent
\begin{equation}
  \makebox[0pt][l]{\hspace*{-\mathindent}\(
    \begin{array}{rcl}
\displaystyle {\Theta}^p_{sk}&=&\displaystyle 4\sum_{n=0}^{+\infty}\frac{1}{(k_BT)^{2n}}[\mathcal{H}_0,p_{sk}]_{2n}\frac{2^{2n+2}-1}{(2n+2)!}B_{2n+2}\\
    &=&\displaystyle p_{sk}-\frac{1}{12(k_BT)^2}[\mathcal{H}_0,[\mathcal{H}_0,p_{sk}]]\displaystyle +\frac{1}{120(k_BT)^4}[\mathcal{H}_0,[\mathcal{H}_0,[\mathcal{H}_0,[\mathcal{H}_0,p_{sk}]]]]...,
    \end{array}
  \)}\hfill
\end{equation}
\noindent
\begin{equation}
  \makebox[0pt][l]{\hspace*{-\mathindent}\(
    \begin{array}{rcl}
\displaystyle \frac{{\Theta}^q_{sk}}{m_k}&=&\displaystyle 4\sum_{n=0}^{+\infty}\frac{1}{(k_BT)^{2n}}[\mathcal{H}_0,\frac{\partial V_0}{\partial r_{sk}}]_{2n}\frac{2^{2n+2}-1}{(2n+2)!}B_{2n+2}\\
    &=&\displaystyle\frac{\partial V_0}{\partial r_{sk}}-\frac{1}{12(k_BT)^2}[\mathcal{H}_0,[\mathcal{H}_0,\frac{\partial V_0}{\partial r_{sk}}]]\displaystyle +\frac{1}{120(k_BT)^4}[\mathcal{H}_0,[\mathcal{H}_0,[\mathcal{H}_0,[\mathcal{H}_0,\frac{\partial V_0}{\partial r_{sk}}]]]]...,
    \end{array}
  \)}\hfill
\end{equation}
where $B_k$ are the Bernoulli numbers $B_2=1/6$, $B_4=-1/30$, $B_6=1/42$,..., and we defined the symbol $[\mathcal{H}_0,\mathcal{A}]_n$ through the recursive relation  $[\mathcal{H}_0,\mathcal{A}]_0=\mathcal{A}$, and $[\mathcal{H}_0,\mathcal{A}]_{n+1}=[\mathcal{H}_0,[\mathcal{H}_0,\mathcal{A}]_n]$ (see Ref. \cite{giordano2025} for details).

\subsection{Application to the quantum harmonic oscillator}
This approach can be applied to the one-dimensional harmonic oscillator (HO) that, besides representing a paradigmatic example of exactly solvable model in quantum mechanics, has a large number of applications ranging from quantum optics to quantum information. The HO is, as usual, defined by the quadratic potential energy $V(x) = \frac{1}{2}m\omega^2q^2$, where $\omega=\sqrt{k/m}$ is the classical angular frequency, and $k$ is the elastic constant. The Hamiltonian of the system is $\mathcal{H}_0=\frac{1}{2m}p^2+\frac{1}{2}m\omega^2q^2$,  whose energy levels are given by $\mathcal{H}_0\varphi_{n}=E_n\varphi_{n}$ with 
$   E_n=\hbar \omega\left(n+\frac{1}{2}\right),\,\,\,\, n\ge 0$;
the HO eigenfunctions can be obtained as
$ \varphi_{n}(q)={\frac {1}{\sqrt {2^{n}\,n!}}}\left({\frac {m\omega }{\pi \hbar }}\right)^{1/4}e^{-{\frac {m\omega }{2\hbar }q^{2}}}H_{n}\left({\sqrt {\frac {m\omega }{\hbar }}}q\right)$,
 where $H_n(z)$ are the Hermite polynomials.
By calculating the matrices associated to the operators $q$ and $p$ as $q_{nm}=\langle \varphi_n(q)\vert x\varphi_m(q) \rangle$ and $p_{nm}=-i\hbar\langle \varphi_n(q)\vert \frac{d}{dq}\varphi_m(q) \rangle$, the following result is obtained:
\begin{equation}
 q_{nm}=\sqrt{\frac{\hbar}{2m\omega}}\left(\delta_{n+1,m}\sqrt{n+1}+\delta_{n,m+1}\sqrt{n}\right), 
\end{equation}
and 
\begin{equation}
 p_{nm}=-i\sqrt{\frac{m\omega\hbar}{2}}\left(\delta_{n+1,m}\sqrt{n+1}-\delta_{n,m+1}\sqrt{n}\right), 
\end{equation}
 for $n\ge 0$ and $m\ge 0$.
Therefore, the corresponding HO friction operators, defined in Eqs. (\ref{raprp}) and (\ref{raprq}), are given by:
\begin{eqnarray}
    \Theta^p=p\frac{\tanh\left(\frac{\hbar \omega}{2k_B T}\right)}{\frac{\hbar \omega}{2k_B T}},\,\,\,\,\, \Theta^q=m^2\omega^2 q\frac{\tanh\left(\frac{\hbar \omega}{2k_B T}\right)}{\frac{\hbar \omega}{2k_B T}}.
    \label{friop}
\end{eqnarray}
We observe that $\Theta^p$ is proportional to $p$, while $\Theta^q$ is proportional to $q$. 
These premises allow us to write the evolution of the density matrix in the form
\begin{eqnarray}
\label{har}
    \frac{d\varrho}{dt}=\frac{1}{i\hbar}\left[\mathcal{H}_0,\varrho\right]&-&\frac{k_B T \beta_p m}{\hbar^2}\left[q,\left[q,\varrho\right]\right]+\frac{\beta_p}{2i\hbar}\frac{\tanh\left(\frac{\hbar \omega}{2k_B T}\right)}{\frac{\hbar \omega}{2k_B T}}\left[q,p\varrho+\varrho p\right]\\
    \nonumber
    &-&\frac{k_B T \beta_q m}{\hbar^2}\left[p,\left[p,\varrho\right]\right]-\frac{ m^2 \omega^2 \beta_q}{2i\hbar}\frac{\tanh\left(\frac{\hbar \omega}{2k_B T}\right)}{\frac{\hbar \omega}{2k_B T}}\left[p,q\varrho+\varrho q\right].
\end{eqnarray}
There are two important particular cases of this equation: (i) when $\beta_q=0$, we obtain the equation discussed in Ref. \cite{giordano2025}, corresponding to the quantization of the standard Langevin equation; and (ii) if $\beta_q=0$, and $\hbar \omega \ll k_B T$, the Caldeira-Legget equation is obtained, with the ratio ${\tanh\left(\frac{\hbar \omega}{2k_B T}\right)}/{\frac{\hbar \omega}{2k_B T}}\simeq 1$ \cite{caldeira1981,caldeira1983,caldeira1983bis}. 

Starting from the general form stated in Eq. (\ref{har}), we now search for the conditions under which this equation is in Lindblad form.

The most general Lindblad master equation \cite{lindblad1974,lindblad1975,gorini1976,lindblad1976,pascazio2017}, which is Markovian, trace-preserving, and completely positive for any initial condition, can be written in the form
\begin{eqnarray}
    \frac{d\varrho}{dt}=\frac{1}{i\hbar}\left[\mathcal{H}_0,\varrho\right]+\sum_i\gamma_i \left[\mathcal{L}_i\varrho \mathcal{L}_i^\dagger-\frac{1}{2}\left(\mathcal{L}_i^\dagger\mathcal{L}_i\varrho+\varrho\mathcal{L}_i^\dagger\mathcal{L}_i\right)\right],
\end{eqnarray}
where $\gamma_i\ge 0$ are non-negative real coefficients, and $\mathcal{L}_i$ are arbitrary operators (the symbol $\dagger$ means ``adjoint operator''). 
In Refs. \cite{dekker1984,sandulescu1987}, by starting from Heisenberg's uncertainty principle, or by using only two Lindblad operators $\mathcal{L}_i=a_ip+b_iq$, $i=1,2$, the authors prove that the equation of the general mathematical form :
\begin{eqnarray}
\label{hargen}
    \frac{d\varrho}{dt}=\frac{1}{i\hbar}\left[\mathcal{H}_0,\varrho\right]&-&\frac{\mathcal{D}_{pp}}{\hbar^2}\left[q,\left[q,\varrho\right]\right]+\frac{\lambda+\mu}{2i\hbar}\left[q,p\varrho+\varrho p\right]+\frac{\mathcal{D}_{pq}}{\hbar^2}\left[q,\left[p,\varrho\right]\right]\\
    \nonumber
    &-&\frac{\mathcal{D}_{qq}}{\hbar^2}\left[p,\left[p,\varrho\right]\right]-\frac{\lambda-\mu}{2i\hbar}\left[p,q\varrho+\varrho q\right]+\frac{\mathcal{D}_{qp}}{\hbar^2}\left[p,\left[q,\varrho\right]\right],
\end{eqnarray}
where $\mathcal{H}_0$ represents the HO and $\mathcal{D}_{qp}=\mathcal{D}_{pq}$, is in the Lindblad form if and only if the following conditions are satisfied:
\begin{eqnarray}
\label{cond1}
    \mathcal{D}_{pp}>0,\\
    \label{cond2}
    \mathcal{D}_{qq}>0,\\
    \label{cond3}
    \mathcal{D}_{pp}\mathcal{D}_{qq}-\mathcal{D}_{pq}^2\ge \frac{\lambda^2\hbar^2}{4}.
\end{eqnarray}
By comparing Eq. (\ref{har}) with Eq. (\ref{hargen}), we see that in our formalism $\mathcal{D}_{qp}=0$ and
\begin{eqnarray}
\label{par1}
    \,\,\,\,\,\,\,\mathcal{D}_{pp}=k_B T \beta_p m,\,\,\,\,\,\,\,\,\,\,\,\,\,\,\,\,\,\,\,\,\,
    \mathcal{D}_{qq}=k_B T \beta_q m,\\
    \label{par2}
    \lambda+\mu=\beta_p\frac{\tanh \xi}{\xi},\,\,\,\,\,\,\,\,\,\,\,
    \lambda-\mu= m^2 \omega^2 \beta_q\frac{\tanh \xi}{\xi},
\end{eqnarray}
where we have defined $\xi=\frac{\hbar \omega}{2k_B T}$ to compact the notation.
First of all, since both $\mathcal{D}_{pp}$ and $\mathcal{D}_{qq}$ must be strictly positive, see Eqs. (\ref{cond1}) and (\ref{cond2}), we obtain from Eq. (\ref{par1}) that both $\beta_p$ and $\beta_q$ must be strictly positive. This definitely demonstrates that in the classical Hamilton-Langevin equation, Eqs. (\ref{ham}) and (\ref{hambis}), the friction and noise terms must be included in both equations so that canonical quantization yields the correct results (completely positive evolution). 

 \begin{figure}[t!]
		\centering
        \includegraphics[width=.52\textwidth]{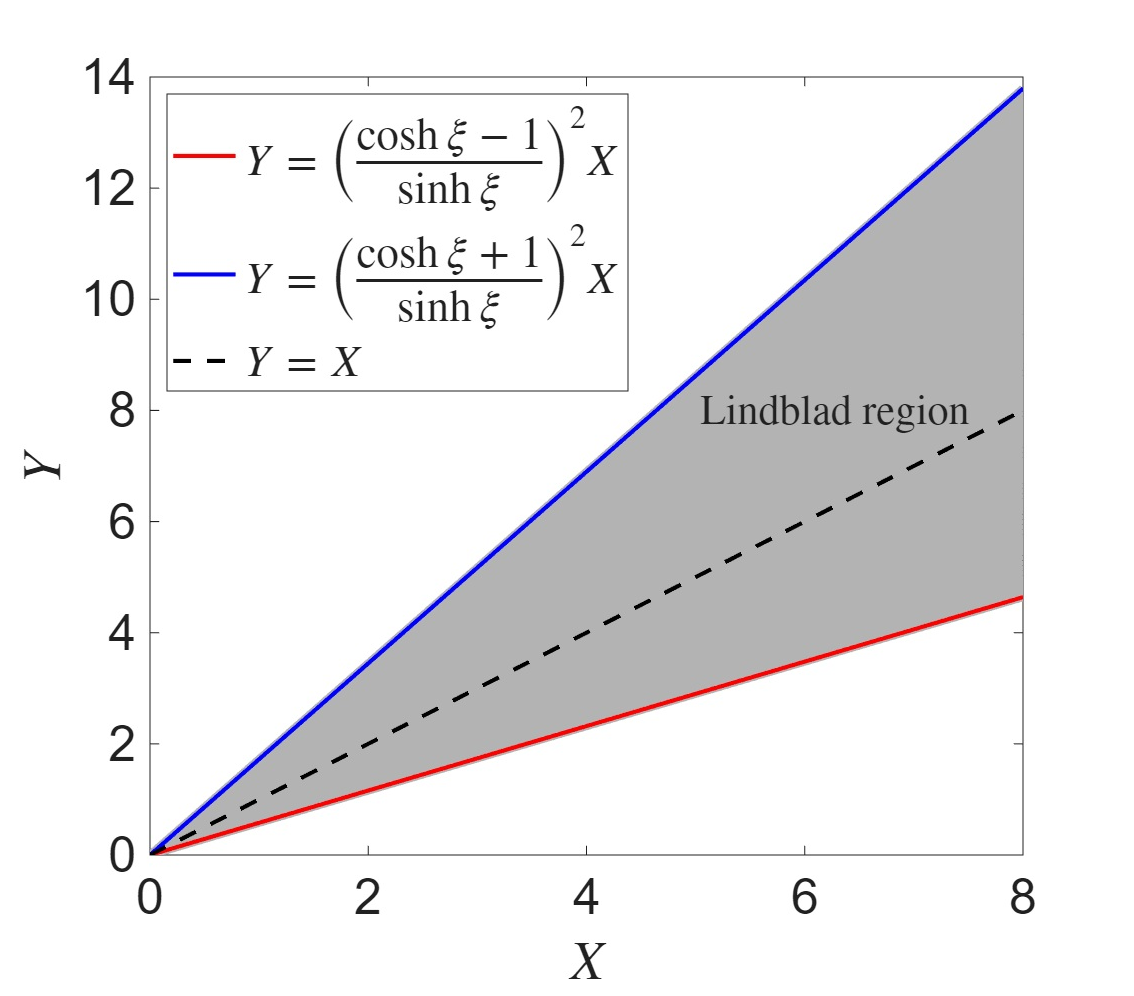}
	\includegraphics[width=.47\textwidth]{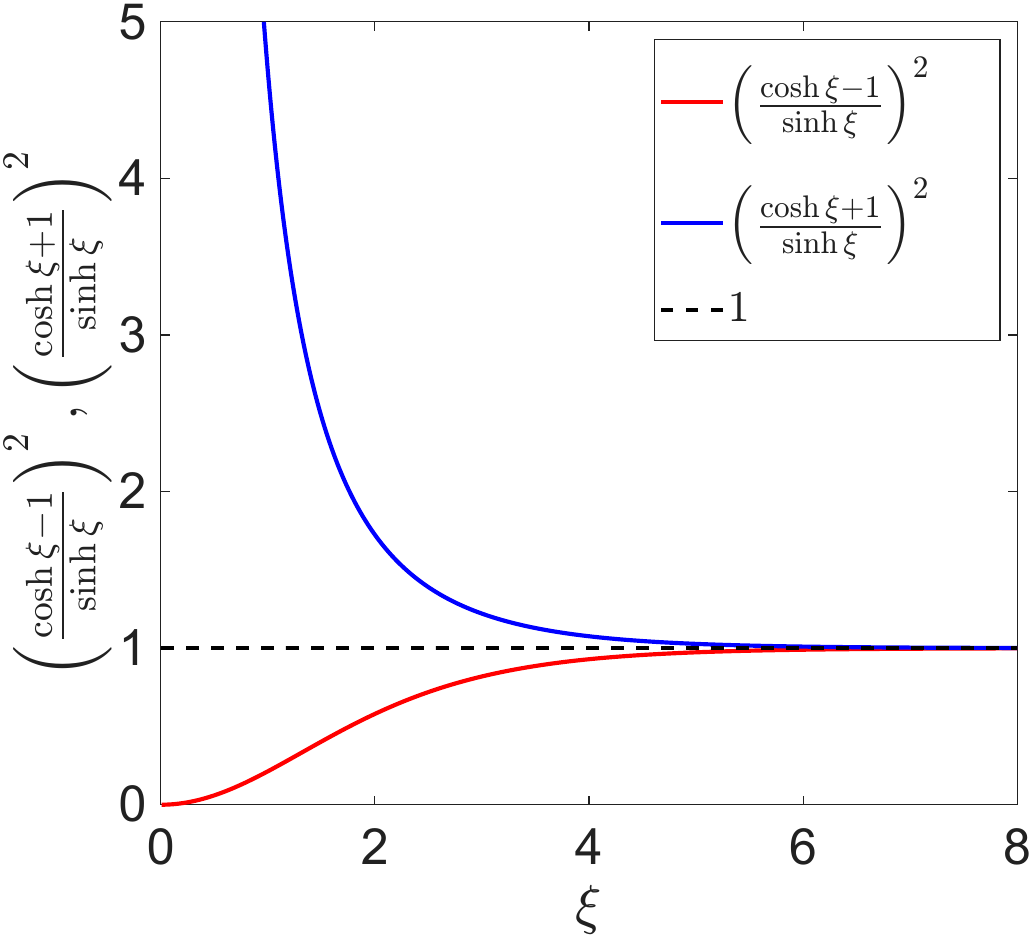}
	\caption{Left: Plot of the straight lines $Y= \left(\frac{\cosh \xi - 1}{\sinh\xi}\right)^2X$, and $Y=\left(\frac{\cosh \xi + 1}{\sinh\xi}\right)^2X$, for $\xi=2$, delimiting the Lindblad region $(Y=\beta_p,X=m^2\omega^2\beta_q)$. The bisector of the first quadrant is also shown for completeness.  Right: Behavior of the functions $\left(\frac{\cosh \xi - 1}{\sinh\xi}\right)^2$, and $\left(\frac{\cosh \xi + 1}{\sinh\xi}\right)^2$, versus $\xi$. For large values of $\xi$, both converge to 1.}
	\label{limits}	
\end{figure}

We also have to study the consequences of the constraint in Eq.(\ref{cond3}). By defining $x=\beta_p/(m^2\omega^2\beta_q)$, this relation can be rewritten as
\begin{eqnarray}
    4x\ge (x+1)^2\tanh^2 \xi\Rightarrow x^2\tanh^2 \xi+2x(\tanh^2 \xi-2)+\tanh^2 \xi\leq 0.
    \label{ine}
\end{eqnarray}
It corresponds to a second-degree inequality with discriminant $\Delta=16(1-\tanh^2 \xi)$, and therefore the two solutions of the associated second-degree equation can be found as
\begin{eqnarray}
\label{sols}
    x_{1,2}=\left(\frac{\cosh \xi\pm 1}{\sinh\xi}\right)^2,
\end{eqnarray}
where we used the property $1-\tanh^2 \xi=1/(\cosh^2\xi)$. 
The values of $x$ that satisfy the inequality in Eq. (\ref{ine}) are in the interval $x_1<x<x_2$, which is written more explicitly as
\begin{eqnarray}
  \left(\frac{\cosh \xi - 1}{\sinh\xi}\right)^2  \leq \frac{\beta_p}{m^2\omega^2\beta_q} \leq \left(\frac{\cosh \xi + 1}{\sinh\xi}\right)^2,
  \label{inte1}
\end{eqnarray}
or, equivalently, as
\begin{eqnarray}
  -1  \leq \frac{1}{m\omega}\sqrt{\frac{\beta_p}{\beta_q}}\sinh\left(\frac{\hbar \omega}{2k_B T}\right)-\cosh\left(\frac{\hbar \omega}{2k_B T}\right) \leq +1.
  \label{inte2}
\end{eqnarray}
This relationship must always be satisfied by the parameters $\beta_p>0$ and $\beta_q>0$, so that the evolution is completely positive. From the geometrical point of view, on the plane $(Y=\beta_p,X=m^2\omega^2\beta_q)$, the region of points that satisfy the inequality corresponds to the angle in the first quadrant between the straight lines $Y= \left(\frac{\cosh \xi - 1}{\sinh\xi}\right)^2X$, and $Y=\left(\frac{\cosh \xi + 1}{\sinh\xi}\right)^2X$. 
The first straight line is always below the bisector of the first quadrant, and the second is always above it, see Fig. \ref{limits}. 
Therefore, a straightforward choice is to take $Y=X$, or $\beta_p=m^2\omega^2\beta_q$, corresponding to $\mu=0$, see Eq. (\ref{par2}). Another possibility is to take $\frac{\beta_p}{m^2\omega^2\beta_q}=\coth^2\xi$, corresponding to the average value of the endpoints of the interval in Eq. (\ref{inte2}).

We underline that, although in the classical case the parameters $\beta_p$ and $\beta_q$ are fully arbitrary, in the quantum counterpart they are subjected to the discussed constraint in order to preserve complete positivity. 

\subsection{Relation with the quantum optical master equation }
In this section, we investigate the connection between the results obtained about the dissipative quantum oscillator within the proposed framework and those based on the Quantum Optical Master Equation (QOME) \cite{englert,gardiner,petruccione}. 
The QMOE, its derivation, approximations, and validity have been widely investigated due to the large spectrum of applications, ranging from quantum optics to quantum information and computation.

In the regime of so-called rotating wave approximation and neglecting Lamb and Stark shifts \cite{englert,gardiner,petruccione}, the QOME for a damped harmonic oscillator (for instance, describing the damping of an electromagnetic field mode inside a cavity) reads
\begin{eqnarray}
\label{QOME}
    \frac{d\varrho}{dt}=-i\omega\left[a^\dagger a,\varrho\right]&+&\frac{\gamma_0}{2} (\bar{n}+1)\left\{2a\varrho a^\dagger- a^\dagger a \varrho-\varrho a^\dagger a\right\}\\&+&\frac{\gamma_0}{2} \bar{n}\left\{2a^\dagger\varrho a- a a^\dagger \varrho-\varrho a a^\dagger\right\},\nonumber
\end{eqnarray}
where the evolution of the density operator is described in terms of the creation and annihilation operators $a^\dagger$ and $a$, respectively. 
While the first term represents the Hamiltonian evolution, the second and third Lindblad terms describe spontaneous emission, thermally induced emission, and absorption processes.
They are described by  the rate $\gamma_0$ (damping of the cavity mode), and by the average number of quanta in the thermal bath in a mode with frequency $\omega$, defined as
\begin{equation}
\bar{n}=\frac{1}{e^{\frac{\hbar \omega}{k_B T}}-1}.
\end{equation}
The relation between these operators and the canonical operators $q$, $p$ is given by
\begin{equation}
a\dagger=\sqrt{\frac{m\omega}{2\hbar}}q-i\sqrt{\frac{1}{2m\hbar\omega}}p,\quad a=\sqrt{\frac{m\omega}{2\hbar}}q+i\sqrt{\frac{1}{2m\hbar\omega}}p,
\label{eq:aadagger}
\end{equation}
satisfying the commutation relations
\begin{equation}
[a,a^\dagger]=1,\quad [a,a]=[a^\dagger,a^\dagger]=0.
\end{equation}
By substituting Eq. (\ref{eq:aadagger}) into the QOME in Eq. (\ref{QOME}), and comparing the results with Eq. (\ref{har}), we can find the the conditions among the coefficients $\beta_p$, $\beta_q$ and the couple $(\gamma_0,\bar{n})$ such that the two equations coincide. After some algebra, we find that the following relation must be verified
\begin{equation}
\beta_p=m^2\omega^2\beta_q=\frac{\gamma_0}{2}\frac{\hbar\omega}{2k_BT}(2\bar{n}+1)=\frac{\gamma_0}{2}\frac{\frac{\hbar\omega}{2k_BT}}{\tanh\left(\frac{\hbar \omega}{2k_B T}\right)}.
\end{equation}
We notice that this result is in perfect agreement with what we found in the previous subsection about the conditions for the Lindblad structure of Eq. (\ref{har}). 
In particular, Eq. (\ref{har}) coincides with the QOME in Eq. (\ref{QOME}) for the choice $\mu=0$ in Eq. (\ref{par2}).
Importantly, it means that the QOME in Eq. (\ref{QOME}) is always represented by a point lying on the bisector represented in Fig. \ref{limits}.
Thus, the results obtained for the damped harmonic oscillator in the quantum framework, described in the previous section, are fully coherent with the outcomes used in the literature based on the QOME.
{  We also note that the cavity model introduced in Ref. \cite{coppola2025} exhibits a clear analogy with our approach based on symmetric friction and noise in the Langevin equations, and is indeed consistent with the optical master equation.}

\section{Second approach: non-Hermitian friction operators}
\label{nonherm}

We describe an alternative procedure to perform the canonical quantization of Eq. (\ref{poisson}). This approach has been recently proposed in the literature in the case with $\beta_q=0$\cite{oliveira2016,oliveira2023,oliveira2024}, and is generalized here for arbitrary values of $\beta_p$ and $\beta_q$ to solve the complete-positivity issue. 

As before, the terms on the third and fifth lines of Eq. (\ref{poisson}),  containing double Poisson brackets, are easily converted to the quantum case using a double commutator.
Instead, the classical terms $m_i\frac{\partial \mathcal{H}_0}{\partial  p_{xi}} W$, $m_i\frac{\partial \mathcal{H}_0}{\partial  p_{yi}} W$, and $m_i\frac{\partial \mathcal{H}_0}{\partial  p_{zi}} W$ can be converted into Hermitian operators by introducing the quantities $\frac{1}{2}(\Xi^{p\dagger}_{xi}{\varrho}+{\varrho}\Xi^{p}_{xi})$, $\frac{1}{2}(\Xi^{p\dagger}_{yi}{\varrho}+{\varrho}\Xi^{p}_{yi})$, and $\frac{1}{2}(\Xi^{p\dagger}_{zi}{\varrho}+{\varrho}\Xi^{p}_{zi})$, where $\Xi^{p}_{xi}$, $\Xi^{p}_{yi}$, and $\Xi^{p}_{zi}$ are non-Hermitian operators to be determined $\forall i=1..N$ (the symbol $\dagger$ means ``adjoint operator''). 
Similarly, the classical terms $m_i\frac{\partial \mathcal{H}_0}{\partial r_{xi}} W$, $m_i\frac{\partial \mathcal{H}_0}{\partial r_{yi}} W$, and $m_i\frac{\partial \mathcal{H}_0}{\partial r_{zi}} W$ are converted into Hermitian operators by introducing the quantities   $\frac{1}{2}(\Xi^{q\dagger}_{xi}{\varrho}+{\varrho}\Xi^{q}_{xi})$, $\frac{1}{2}(\Xi^{q\dagger}_{yi}{\varrho}+{\varrho}\Xi^{q}_{yi})$, and $\frac{1}{2}(\Xi^{q\dagger}_{zi}{\varrho}+{\varrho}\Xi^{q}_{zi})$, where $\Xi^{q}_{xi}$, $\Xi^{q}_{yi}$, and $\Xi^{q}_{zi}$ are non-Hermitian operators to be determined ($\forall i=1..N$).
This choice leads to some simplifications in the calculations, however it must be kept in mind that non-Hermitian friction operators can not be directly associated with real physical observables.

With these premises, the complete equation, analogous of Eq.(\ref{mastereq1}) above, is eventually obtained:
\noindent
\begin{equation}
\label{maindiffforcesbis}
  \makebox[0pt][l]{\hspace*{-\mathindent}\(
    \begin{array}{rcl}
\displaystyle   \frac{d\varrho}{dt}&=&\displaystyle\frac{1}{i\hbar}\left[\mathcal{H}_0,\varrho\right]-\frac{1}{i\hbar}\sum_{k=1}^{N}\sum_{s=x,y,z}f_{sk}\left[r_{sk},\varrho\right]\\
    &-&\displaystyle\frac{k_B T \beta_p}{\hbar^2}\sum_{k=1}^{N}m_k\left(\sum_{s=x,y,z}\left[r_{sk},\left[r_{sk},\varrho\right]\right]\right)+\frac{\beta_p}{2i\hbar}\sum_{k=1}^{N}\left(\sum_{s=x,y,z}\left[r_{sk},\Xi^{p\dagger}_{sk}\varrho+\varrho\Xi^{p}_{sk}\right]\right)\\
    &-&\displaystyle\frac{k_B T \beta_q}{\hbar^2}\sum_{k=1}^{N}m_k\left(\sum_{s=x,y,z}\left[p_{sk},\left[p_{sk},\varrho\right]\right]\right)-\frac{\beta_q}{2i\hbar}\sum_{k=1}^{N}\left(\sum_{s=x,y,z}\left[p_{sk},\Xi^{q\dagger}_{sk}\varrho+\varrho\Xi^{q}_{sk}\right]\right),
    \end{array}
  \)}\hfill
\label{mastereq2}
\end{equation}
which represents the new version of the quantum master equation.
The friction operators $\Xi^{p,q}_{sk}$ must be obtained by imposing the asymptotic canonical quantum distribution stated in Eq. (\ref{boltz}).
Therefore, the following relations must be satisfied:
\begin{eqnarray}
    \frac{2im_k k_B T}{\hbar}\left[r_{sk},e^{-\frac{\mathcal{H}_0}{k_BT}}\right]&=&\Xi^{p\dagger}_{sk}e^{-\frac{\mathcal{H}_0}{k_BT}}+e^{-\frac{\mathcal{H}_0}{k_BT}}\Xi^{p}_{sk},
    \label{mainequationpbis}
\\
   -\frac{2im_k k_B T}{\hbar}\left[p_{sk},e^{-\frac{\mathcal{H}_0}{k_BT}}\right]&=&\Xi^{q\dagger}_{sk}e^{-\frac{\mathcal{H}_0}{k_BT}}+e^{-\frac{\mathcal{H}_0}{k_BT}}\Xi^{q}_{sk},
    \label{mainequationqbis}
\end{eqnarray}
for $s=x,y,z$, and $k=1,...,N$.
The solutions can be easily obtained in the form
\noindent
\begin{equation}
\label{fribis}
  \makebox[0pt][l]{\hspace*{-\mathindent}\(
    \begin{array}{rcl}
    \displaystyle\Xi^{p}_{sk}&=&\displaystyle\frac{im_k k_B T}{\hbar}\left(e^{+\frac{\mathcal{H}_0}{k_BT}}r_{sk}e^{-\frac{\mathcal{H}_0}{k_BT}}-r_{sk}\right),\,\,\,\,\,\,\,\,\,\,\,\,  \Xi^{p\dagger}_{sk}=-\frac{im_k k_B T}{\hbar}\left(e^{-\frac{\mathcal{H}_0}{k_BT}}r_{sk}e^{+\frac{\mathcal{H}_0}{k_BT}}-r_{sk}\right),\\
    \displaystyle\Xi^{q}_{sk}&=&\displaystyle-\frac{im_k k_B T}{\hbar}\left(e^{+\frac{\mathcal{H}_0}{k_BT}}p_{sk}e^{-\frac{\mathcal{H}_0}{k_BT}}-p_{sk}\right),\,\,\,\,  \Xi^{q\dagger}_{sk}=\frac{im_k k_B T}{\hbar}\left(e^{-\frac{\mathcal{H}_0}{k_BT}}p_{sk}e^{+\frac{\mathcal{H}_0}{k_BT}}-p_{sk}\right),
    \end{array}
  \)}\hfill
\end{equation}
where we also show the adjoint operators.

Now, we remember that any non-Hermitian operator $\mathcal{A}\neq\mathcal{A}^{\dagger}$ can be always written in the form $\mathcal{A}=\mathcal{R}+i\mathcal{I}$, where $\mathcal{R}$ and $\mathcal{I}$ are Hermitian operators, with  $\mathcal{R}=\mathcal{R}^{\dagger}$ and $\mathcal{I}=\mathcal{I}^{\dagger}$. The relations $\mathcal{R}=\frac{1}{2}(\mathcal{A}+\mathcal{A}^{\dagger})$ and $\mathcal{I}=\frac{1}{2i}(\mathcal{A}-\mathcal{A}^{\dagger})$ can be easily proved. The first friction operator $\Xi^{p}_{sk}$ can be therefore decomposed in the form $\Xi^{p}_{sk}=\mathcal{A}^{p}_{sk}+i\mathcal{B}^{p}_{sk}$, where:
\begin{eqnarray}
    \mathcal{A}^{p}_{sk}&=&\frac{im_k k_B T}{2\hbar}\left(e^{+\frac{\mathcal{H}_0}{k_BT}}r_{sk}e^{-\frac{\mathcal{H}_0}{k_BT}}-e^{-\frac{\mathcal{H}_0}{k_BT}}r_{sk}e^{+\frac{\mathcal{H}_0}{k_BT}}\right),\\
    \mathcal{B}^{p}_{sk}&=&\frac{m_k k_B T}{2\hbar}\left(e^{+\frac{\mathcal{H}_0}{k_BT}}r_{sk}e^{-\frac{\mathcal{H}_0}{k_BT}}+e^{-\frac{\mathcal{H}_0}{k_BT}}r_{sk}e^{+\frac{\mathcal{H}_0}{k_BT}}-2r_{sk}\right).
\end{eqnarray}
Similarly, the second friction operator $\Xi^{q}_{sk}$ can be decomposed in the form $\Xi^{q}_{sk}=\mathcal{A}^{q}_{sk}+i\mathcal{B}^{q}_{sk}$, where
\begin{eqnarray}
    \mathcal{A}^{q}_{sk}&=&-\frac{im_k k_B T}{2\hbar}\left(e^{+\frac{\mathcal{H}_0}{k_BT}}p_{sk}e^{-\frac{\mathcal{H}_0}{k_BT}}-e^{-\frac{\mathcal{H}_0}{k_BT}}p_{sk}e^{+\frac{\mathcal{H}_0}{k_BT}}\right),\\
    \mathcal{B}^{q}_{sk}&=&-\frac{m_k k_B T}{2\hbar}\left(e^{+\frac{\mathcal{H}_0}{k_BT}}p_{sk}e^{-\frac{\mathcal{H}_0}{k_BT}}+e^{-\frac{\mathcal{H}_0}{k_BT}}p_{sk}e^{+\frac{\mathcal{H}_0}{k_BT}}-2p_{sk}\right).
\end{eqnarray}
These expressions are particularly useful when the operators are represented in the energy basis induced by the eigenvector equation $\mathcal{H}_0\varphi_n(\vec{q})=E_n\varphi_n(\vec{q})$. Under this assumption, the representation of the friction operators is obtained in the following form:
\noindent
\begin{equation}
\label{fritris}
  \makebox[0pt][l]{\hspace*{-\mathindent}\(
    \begin{array}{rcl}
    \displaystyle\mathcal{A}^{p}_{sk,\ell j}&=&\displaystyle p_{sk,\ell j}\frac{\sinh\left(\frac{E_{\ell}-E_j}{k_B T}\right)}{\frac{E_{\ell}-E_j}{k_B T}},\,\,\,\,\,\,\,\,\,\,\,\,\,\,\,\,\,\,\,\,\,\,\,\,\,\,\,\,\,\,\,\,\,\,\,\,
    \mathcal{B}^{p}_{sk,\ell j}=\frac{m_k k_B T}{\hbar}r_{sk,\ell j}\left[\cosh\left(\frac{E_{\ell}-E_j}{k_B T}\right)-1\right],\\
    \displaystyle\mathcal{A}^{q}_{sk,\ell j}&=&\displaystyle m_k\left(\frac{\partial V_0}{\partial r_{sk}}\right)_{\ell j}\frac{\sinh\left(\frac{E_{\ell}-E_j}{k_B T}\right)}{\frac{E_{\ell}-E_j}{k_B T}},\,\,\,\,\,
    \mathcal{B}^{p}_{sk,\ell j}=-\frac{m_k k_B T}{\hbar}p_{sk,\ell j}\left[\cosh\left(\frac{E_{\ell}-E_j}{k_B T}\right)-1\right].
  \end{array}
  \)}\hfill
\end{equation}
These expressions are important because they show once again that in the classical limit ($\hbar\rightarrow 0$, or $E_{\ell}-E_j\rightarrow 0$), we find the limiting values   $\Xi^{p}_{sk}=\mathcal{A}^{p}_{sk}+i\mathcal{B}^{p}_{sk}\rightarrow p_{sk}$ and $\Xi^{q}_{sk}=\mathcal{A}^{q}_{sk}+i\mathcal{B}^{q}_{sk}\rightarrow m_k\frac{\partial V_0}{\partial r_{sk}}$, as discussed for the classical Fokker-Planck equation.
As already mentioned, the analytical expressions of the friction operators are quite manageable in this case, but they lose the Hermitian character that is typical of genuine physical observables.

\subsection{Application to the quantum harmonic oscillator}

We can apply this formalism to the one-dimensional HO to study the deviations from the previous case.
As before, we consider the Hamiltonian of the system in the form $\mathcal{H}_0=\frac{1}{2m}p^2+\frac{1}{2}m\omega^2q^2$, and we obtain the explicit form of the friction operators as:
\begin{eqnarray}
\label{gggp}
    \Xi^p=\frac{\sinh\left(\frac{\hbar\omega}{k_B T}\right)}{\frac{\hbar\omega}{k_B T}}p+im\omega\frac{\cosh\left(\frac{\hbar\omega}{k_B T}\right)-1}{\frac{\hbar\omega}{k_B T}}q,\\
    \label{gggq}
    \Xi^q=m^2\omega^2\frac{\sinh\left(\frac{\hbar\omega}{k_B T}\right)}{\frac{\hbar\omega}{k_B T}}q-im\omega\frac{\cosh\left(\frac{\hbar\omega}{k_B T}\right)-1}{\frac{\hbar\omega}{k_B T}}p.
\end{eqnarray}

In this case, the friction operators are given by a linear combination of the operators $p$ and $q$. By comparing them with Eq. (\ref{friop}), it is observed that in any case when $\frac{\hbar\omega}{k_B T}\rightarrow 0$ the two operators converge to $p$ and $m^2\omega^2 q$, respectively.
The obtained operators in Eqs.(\ref{gggp}) and (\ref{gggq}) can be inserted into the quantum master equation stated in Eq.(\ref{maindiffforcesbis}), to eventually obtain
\begin{eqnarray}
\label{harbis}
    \frac{d\varrho}{dt}&=&\frac{1}{i\hbar}\left[\mathcal{H}_0,\varrho\right]-\frac{k_B T \beta_p m}{\hbar^2}\left[q,\left[q,\varrho\right]\right]+\frac{\beta_p}{2i\hbar}\frac{\sinh\left(\frac{\hbar \omega}{k_B T}\right)}{\frac{\hbar \omega}{k_B T}}\left[q,p\varrho+\varrho p\right]\\
    \nonumber
    &-&\frac{k_B T \beta_q m}{\hbar^2}\left[p,\left[p,\varrho\right]\right]-\frac{ m^2 \omega^2 \beta_q}{2i\hbar}\frac{\sinh\left(\frac{\hbar \omega}{k_B T}\right)}{\frac{\hbar \omega}{k_B T}}\left[p,q\varrho+\varrho q\right]\\
    \nonumber
    &-&\frac{m\omega\beta_p}{2\hbar}\frac{\cosh\left(\frac{\hbar\omega}{k_B T}\right)-1}{\frac{\hbar\omega}{k_B T}}\left[q,\left[q,\varrho\right]\right]-\frac{m\omega\beta_q}{2\hbar}\frac{\cosh\left(\frac{\hbar\omega}{k_B T}\right)-1}{\frac{\hbar\omega}{k_B T}}\left[p,\left[p,\varrho\right]\right].
\end{eqnarray}
This equation can be compared with Eq.(\ref{hargen}), and we can identify the crucial parameters for verifying its Lindblad character. 
First of all, we have that  $\mathcal{D}_{qp}=0$, and moreover we get
\begin{eqnarray}
\label{par1bis}
    {\mathcal{D}_{pp}}&=&{k_B T \beta_p m}+\frac{m\hbar\omega\beta_p}{2}\frac{\cosh\eta-1}{\eta},\\
    \label{par2bis}
    {\mathcal{D}_{qq}}&=&{k_B T \beta_q m}+\frac{m\hbar\omega\beta_q}{2}\frac{\cosh\eta-1}{\eta},\\
    \label{par3bis}
    \lambda+\mu&=&\beta_p\frac{\sinh \eta}{\eta},\,\,\,\,\,\,\,\,\,\,\,
    \lambda-\mu= m^2 \omega^2 \beta_q\frac{\sinh \eta}{\eta},
\end{eqnarray}
where we have defined $\eta=\frac{\hbar \omega}{k_B T}$ to compact the notation.
As in the previous case, since both $\mathcal{D}_{pp}$ and $\mathcal{D}_{qq}$ must be strictly positive, see Eqs. (\ref{cond1}) and (\ref{cond2}), we obtain from Eqs. (\ref{par1bis}) and (\ref{par2bis}) that both $\beta_p$ and $\beta_q$ must be strictly positive. Again, this proves that we must include the friction and noise terms in both Hamilton-Langevin equations, see Eqs. (\ref{ham}) and (\ref{hambis}). 
We now investigate the condition in Eq.(\ref{cond3}). By defining again $x=\beta_p/(m^2\omega^2\beta_q)$, this relation can be rewritten as
\begin{eqnarray}
    4x\left(1+\cosh\eta\right)^2\ge (x+1)^2\sinh^2 \eta,
    \label{inebis}
\end{eqnarray}
which is equivalent to
\begin{eqnarray}
    x^2\sinh^2 \eta-x\left(8+8\cosh\eta+2\sinh^2\eta\right)+\sinh^2\eta\le 0.
    \label{inetris}
\end{eqnarray}
This second-degree inequality leads to a discriminant $\Delta=32(1+\cosh\eta)^3=256\cosh^6\left(\frac{\eta}{2}\right)=256\cosh^6\xi$, where $\xi=\frac{\hbar \omega}{2k_B T}$, as before.
Hence, we determine the two solutions as
\begin{eqnarray}
\label{solsbis}
    x_{1,2}=\left[\frac{\cosh\left(\frac{\eta}{2}\right)\pm 1}{\sinh\left(\frac{\eta}{2}\right)}\right]^2=\left(\frac{\cosh \xi\pm 1}{\sinh\xi}\right)^2,
\end{eqnarray}
where we used the standard relations $\cosh\eta=\sinh^2\xi+\cosh^2\xi$, and $\sinh\eta=2\sinh\xi\cosh\xi$.

Importantly, these solutions coincide with those obtained in Eq. (\ref{sols}). Hence, also in this second case, we recover the condition under which the equation attains the Lindblad form, which are precisely the same as derived in the first case,  expressed in Eqs. (\ref{inte1}) or (\ref{inte2}). Remarkably, this condition turns out to be independent of the specific choice of friction operators, suggesting that it could represent a more general, possibly universal result.

\section{Quantum stochastic thermodynamics}
\label{thermo}

In this section, we construct the thermodynamic formulation starting from the evolution equation for the density matrix introduced in Eq. (\ref{maindiffforces}). For the sake of clarity and without loss of generality, we confine our discussion to the case of Hamiltonian friction operators, which are not only analytically tractable but also the most relevant from a physical standpoint. Within this framework, we establish the first and second laws of thermodynamics after introducing the notions of internal energy and entropy associated with the system.
{  The same line of analysis could be extended to the second case with non-Hermitian friction operators; nevertheless, we refrain from presenting it here for conciseness, as it yields results essentially analogous to those described below.}

\subsection{First law of thermodynamics}

To  retrieve the first law of thermodynamics, we firstly introduce the internal energy $\mathcal{E}$ of the system, defined as the expectation value of the Hamiltonian operator, $\mathcal{E}=\mbox{Tr}(\mathcal{H}_0\varrho)$.
The time evolution of this internal energy is: 
\begin{eqnarray}
\frac{d\mathcal{E}}{dt}&=&\frac{d\mbox{Tr}(\mathcal{H}_0\varrho)}{dt}=\mbox{Tr}\left(\mathcal{H}_0\frac{d\varrho}{dt}\right).
\end{eqnarray}
We need to substitute here the
terms coming from Eq. (\ref{maindiffforces}). 
The classical Liouvillian term leads to $
\mbox{Tr}\left(\mathcal{H}_0\left[\mathcal{H}_0,\varrho\right]\right)=0$,
because of the cyclic property of the trace.
Concerning the effect of the external forces, we need to develop $\mbox{Tr}\left(\mathcal{H}_0\left[r_{sk},\varrho\right]\right)=\mbox{Tr}\left(\left[\mathcal{H}_0,r_{sk}\right]\varrho\right)$, where we used again the cyclic property of the trace. We can now use relation $[r_{sk},\mathcal{H}_0]=\frac{i\hbar}{m_k}p_{sk}$, and we get $ 
\mbox{Tr}\left(\mathcal{H}_0\left[r_{sk},\varrho\right]\right)=-i\hbar\mathbb{E}\left\lbrace{p_{sk}}\right\rbrace/m_k$. 
An arbitrary noise term delivers the contribution: 
\begin{eqnarray}
\nonumber
&&\mbox{Tr}\left(\mathcal{H}_0\left[r_{sk},\left[r_{sk},\varrho\right]\right]\right)=\mbox{Tr}\left(\left[\mathcal{H}_0,r_{sk}\right]\left[r_{sk},\varrho\right]\right)\\
&&\quad=-\frac{i\hbar}{m_k}\mbox{Tr}\left(p_{sk}\left[r_{sk},\varrho\right]\right)=-\frac{i\hbar}{m_k}\mbox{Tr}\left(\left[p_{sk},r_{sk}\right]\varrho\right)=-\frac{\hbar^2}{m_k},
\end{eqnarray}
or
\begin{eqnarray}
\nonumber
&&\mbox{Tr}\left(\mathcal{H}_0\left[p_{sk},\left[p_{sk},\varrho\right]\right]\right)=\mbox{Tr}\left(\left[\mathcal{H}_0,p_{sk}\right]\left[p_{sk},\varrho\right]\right)\\
&&\quad=i\hbar\mbox{Tr}\left(\frac{\partial V_0}{\partial r_{sk}}\left[p_{sk},\varrho\right]\right)=i\hbar\mbox{Tr}\left(\left[\frac{\partial V_0}{\partial r_{sk}},p_{sk}\right]\varrho\right)=-{\hbar^2}\mathbb{E}\left\lbrace\frac{\partial^2 V_0}{\partial r_{sk}^2}\right\rbrace.
\end{eqnarray}
Moreover, the arbitrary friction contribution corresponds to the term:
\begin{eqnarray}
\nonumber
&&\mbox{Tr}\left(\mathcal{H}_0\left[r_{sk},\Theta^p_{sk}\varrho+\varrho\Theta^p_{sk}\right]\right)=\mbox{Tr}\left(\left[\mathcal{H}_0,r_{sk}\right]\left(\Theta^p_{sk}\varrho+\varrho\Theta^p_{sk}\right)\right)\\
\nonumber
&&\quad=-\frac{i\hbar}{m_k}\mbox{Tr}\left(p_{sk}\left(\Theta^p_{sk}\varrho+\varrho\Theta^p_{sk}\right)\right)=-\frac{i\hbar}{m_k}\mathbb{E}\left\lbrace{p_{sk}\Theta^p_{sk}+\Theta^p_{sk}p_{sk}}\right\rbrace,
\end{eqnarray}
or
\begin{eqnarray}
\nonumber
&&\mbox{Tr}\left(\mathcal{H}_0\left[p_{sk},\Theta^q_{sk}\varrho+\varrho\Theta^q_{sk}\right]\right)=\mbox{Tr}\left(\left[\mathcal{H}_0,p_{sk}\right]\left(\Theta^q_{sk}\varrho+\varrho\Theta^q_{sk}\right)\right)\\
\nonumber
&&\quad=i\hbar\mbox{Tr}\left(\frac{\partial V_0}{\partial r_{sk}}\left(\Theta^q_{sk}\varrho+\varrho\Theta^q_{sk}\right)\right)=i\hbar\mathbb{E}\left\lbrace\frac{\partial V_0}{\partial r_{sk}}\Theta^q_{sk}+\Theta^q_{sk}\frac{\partial V_0}{\partial r_{sk}}\right\rbrace.
\end{eqnarray}
Summing up all contributions, we obtain the following result:
\begin{eqnarray}
\label{first1}
\frac{d\mathcal{E}}{dt}&=&\sum_{k=1}^{N}\sum_{s=x,y,z}f_{sk}\frac{1}{m_k}\mathbb{E}\left\lbrace{p_{sk}}\right\rbrace\\
\nonumber
&&+2\beta_p\left[\frac{3}{2}N k_B T-\sum_{k=1}^{N}\sum_{s=x,y,z}\frac{1}{2m_k}\mathbb{E}\left\lbrace{\frac{p_{sk}\Theta^p_{sk}+\Theta^p_{sk}p_{sk}}{2}}\right\rbrace\right]\\
\nonumber
&&+\beta_q\left[k_BT\sum_{k=1}^{N}\sum_{s=x,y,z}m_k\mathbb{E}\left\lbrace\frac{\partial^2 V_0}{\partial r_{sk}^2}\right\rbrace-\sum_{k=1}^{N}\sum_{s=x,y,z}\mathbb{E}\left\lbrace\frac{\frac{\partial V_0}{\partial r_{sk}}\Theta^q_{sk}+\Theta^q_{sk}\frac{\partial V_0}{\partial r_{sk}}}{2}\right\rbrace\right]
\end{eqnarray}
This represents the first law of thermodynamics, where we can identify the average rate of work $\frac{d\mathbb{E}\left\lbrace L \right\rbrace}{dt}$ done on the system with the average power $\sum_{k=1}^{N}\vec{f}_{k}\cdot\mathbb{E}\left\lbrace{\vec{p}_{k}}\right\rbrace/m_k$ of the external forces. This term is identical to the one obtained in the classical analysis of the problem.
The second and third terms in Eq. (\ref{first1}) represent the average heat rate $\frac{d\mathbb{E}\left\lbrace Q \right\rbrace}{dt}$ entering the system.
We know that the Hermitian operators $\Theta^p_{sk}$ take the role of classical momenta $p_{sk}$. Similarly, the Hermitian operators $\Theta^q_{sk}$ are the quantum counterpart of the classical terms $m_k\frac{\partial V_0}{\partial r_{sk}}$.
Therefore,  the term $\sum_{k=1}^{N}\sum_{s=x,y,z}\frac{1}{2m_k}\mathbb{E}\left\lbrace{\frac{p_{sk}\Theta^p_{sk}+\Theta^p_{sk}p_{sk}}{2}}\right\rbrace$ represents the quantum modified average kinetic energy of the system, corresponding to the classical term $\sum_{k=1}^{N}\frac{1}{2m_k}\mathbb{E}\left\lbrace{\vec{p}_{k}\cdot\vec{p}_{k}}\right\rbrace$.
Moreover, the quantum term $\sum_{k=1}^{N}\sum_{s=x,y,z}\mathbb{E}\left\lbrace\left(\frac{\partial V_0}{\partial r_{sk}}\Theta^q_{sk}+\Theta^q_{sk}\frac{\partial V_0}{\partial r_{sk}}\right)/2\right\rbrace$, corresponds to the classical contribution $\sum_{k=1}^{N}\sum_{s=x,y,z}m_k\mathbb{E}\left\lbrace\left(\frac{\partial V_0}{\partial r_{sk}}\right)^2\right\rbrace$. Therefore, it can be affirmed that the quantum version of the first law perfectly corresponds to the classical one, stated in Eq. (\ref{1pr}).

It is worth noting that the friction coefficients $\beta_p$ and $\beta_q$ determine the characteristic rate of relaxation toward equilibrium, namely the rate at which equipartition is established.
Indeed, the two bracketed terms multiplying the coefficients $\beta_p$ and $\beta_q$ in Eq. (\ref{first1}) identically vanish at thermodynamic equilibrium. The corresponding identities thus obtained can be interpreted as two distinct formulations of the quantum equipartition theorem.
They can be written as:
\begin{eqnarray}
\label{equi1}
    \frac{1}{2} k_B T=\frac{1}{2m_k}\mathbb{E}\left\lbrace{\frac{p_{sk}\Theta^p_{sk}+\Theta^p_{sk}p_{sk}}{2}}\right\rbrace,\\
    \label{equi2}
    k_BTm_k\mathbb{E}\left\lbrace\frac{\partial^2 V_0}{\partial r_{sk}^2}\right\rbrace=\mathbb{E}\left\lbrace\frac{\frac{\partial V_0}{\partial r_{sk}}\Theta^q_{sk}+\Theta^q_{sk}\frac{\partial V_0}{\partial r_{sk}}}{2}\right\rbrace.
\end{eqnarray}
These two relations represent the quantum version of Eq. (\ref{equic}), and they can be directly proved by using the equilibrium density matrix stated in Eq. (\ref{boltz}). The proof of Eq. (\ref{equi1}) can be found in Ref. \cite{giordano2025}, and the proof of Eq. (\ref{equi2}) can be performed in a very similar way (omitted here for brevity).

\subsection{Second law of thermodynamics}

In this section, we develop a balance equation for the von Neumann entropy of the system defined as follows:
\begin{equation}
    \mathcal{S}=-k_B\mbox{Tr}(\varrho\ln\varrho)=-k_B\mathbb{E}(\ln\varrho).
    \label{qentropy}
\end{equation}
To begin, we calculate the time derivative of entropy:
\begin{equation}
    \frac{d\mathcal{S}}{dt}=-k_B\mbox{Tr}\left(\frac{d\varrho}{dt}\ln\varrho+\varrho\frac{d}{dt}\ln\varrho\right),
    \label{derentro}
\end{equation}
and we remember that $\mbox{Tr}\left(\varrho\frac{d}{dt}\ln\varrho\right)=0$ (see Ref. \cite{giordano2025}). Hence, we obtain the simpler relation:
\begin{equation}
    \frac{d\mathcal{S}}{dt}=-k_B\mbox{Tr}\left(\frac{d\varrho}{dt}\ln\varrho\right).
\end{equation}
The entropy balance can be further elaborated by considering the density matrix evolution given in Eq. (\ref{maindiffforces}).
The Hamiltonian contribution leads to the following term:
\begin{eqnarray}
\mbox{Tr}\left(\left[\mathcal{H}_0,\varrho\right]\ln\varrho\right)=\mbox{Tr}\left(\mathcal{H}_0\varrho\ln\varrho-\varrho\mathcal{H}_0\ln\varrho\right)=\mbox{Tr}\left(\mathcal{H}_0\left[\varrho,\ln\varrho\right]\right)=0,
\end{eqnarray}
where we applied the cyclic property of trace, and recalled that $\varrho$ commutes with $\ln\varrho$.
The contribution of external forces is also zero, in fact:
\begin{eqnarray}
\mbox{Tr}\left(\left[r_{sk},\varrho\right]\ln\varrho\right)=\mbox{Tr}\left(r_{sk}\varrho\ln\varrho-\varrho r_{sk}\ln\varrho\right)=\mbox{Tr}\left(r_{sk}\left[\varrho,\ln\varrho\right]\right)=0.
\end{eqnarray}
To complete the entropy balance, we rewrite Eq. (\ref{maindiffforces}) in the compact form:
\begin{equation}
    \frac{d\varrho}{dt}=\frac{1}{i\hbar}\left[\mathcal{H}_0,\varrho\right]+\mathcal{A}\varrho+\mathcal{R}\varrho=\mathcal{L}\varrho.
    \label{denseq}
\end{equation}
Here, $\mathcal{A}\varrho$ denotes the contribution of the external force, $\mathcal{R}\varrho$ encompasses the friction and noise terms (i.e., the quantum Langevin bath), and $\mathcal{L}\varrho$ represents the sum of all contributions in the evolution equation. Since the Hamiltonian and external force contributions vanish in the entropic balance, we may then write:
\begin{equation}
    \frac{d\mathcal{S}}{dt}=-k_B\mbox{Tr}\left(\mathcal{R}\varrho\ln\varrho\right).
\end{equation}
Similarly, the average heat rate defined in the previous Section can be written as:
\begin{equation}
    \frac{d\mathbb{E}\left\lbrace Q\right\rbrace}{dt}=\mbox{Tr}\left(\mathcal{R}\varrho\mathcal{H}_0\right)=k_B T\mbox{Tr}\left(\mathcal{R}\varrho\frac{\mathcal{H}_0}{k_B T}\right).
\end{equation}
Therefore, we can write the second law of thermodynamics as:
\begin{eqnarray}
    \frac{d\mathcal{S}}{dt}&=&\frac{1}{T}\frac{d\mathbb{E}\left\lbrace Q\right\rbrace}{dt}+\frac{d\mathcal{S}_p}{dt},
\end{eqnarray}
where
\begin{equation}
    \frac{d\mathcal{S}_p}{dt}=-k_B\mbox{Tr}\left[\mathcal{R}\varrho\left(\ln \varrho+\frac{\mathcal{H}_0}{k_B T}\right)\right].
    \label{entrprod}
\end{equation}
The term $\frac{d\mathcal{S}_f}{dt}=\frac{1}{T}\frac{d\mathbb{E}\left\lbrace Q\right\rbrace}{dt}$ represents the entropy flow, that is, the amount of entropy entering the system due to heat exchange. 
The term $\frac{d\mathcal{S}_p}{dt}$ represents the production of entropy, due to the irreversibility of the thermodynamic transformation. It must always be positive for consistency with classical thermodynamics. 

The entropy production term can be simplified by introducing  the asymptotic canonical distribution
$ \varrho_{eq}=\frac{1}{Z_{qu}}e^{-\frac{\mathcal{H}_0}{k_BT}}$. Its logarithm is given by
$\ln\varrho_{eq}=\left(\ln\frac{1}{Z_{qu}}\right)I-\frac{\mathcal{H}_0}{k_B T}$ (see Ref. \cite{giordano2025}).
Since the term $\left(\ln\frac{1}{Z_{qu}}\right)I$ does not contribute to the entropy production, owing to $\mbox{Tr}\left[\mathcal{R}\varrho\right]=0$, the quantity $ \frac{d\mathcal{S}_p}{dt}$ can be expressed in the following form:
\begin{equation}
    \frac{d\mathcal{S}_p}{dt}=k_B\mbox{Tr}\left[\mathcal{R}\varrho\left(\ln \varrho_{eq}-\ln\varrho\right)\right].
    \label{entroprodfin}
\end{equation}
We now demonstrate that the rate of entropy production remains strictly positive throughout the relaxation toward equilibrium. To this end, we consider the system in the absence of any externally applied force, whose dynamics is governed by the evolution equation $\frac{d\varrho}{dt}=\frac{1}{i\hbar}\left[\mathcal{H}_0,\varrho\right]+\mathcal{R}\varrho=\mathcal{L}\varrho$. From Eq. (\ref{entroprodfin}), an equivalent expression for the entropy production can be written as:
\begin{equation}
    \frac{d\mathcal{S}_p}{dt}=k_B\mbox{Tr}\left[\mathcal{L}\varrho\left(\ln \varrho_{eq}-\ln\varrho\right)\right].
    \label{entroprodnew}
\end{equation}
Here, the relaxation operator $\mathcal{R}$ has been replaced by 
$\mathcal{L}$ since $\mbox{Tr}\left(\left[\mathcal{H}_0,\varrho\right]\ln\varrho_{eq}\right)=\mbox{Tr}\left(\left[\mathcal{H}_0,\varrho\right]\ln\varrho\right)=0$.
We note that the evolution equation $\frac{d\varrho}{dt}=\mathcal{L}\varrho$ admits the formal solution $\varrho(t)=e^{\mathcal{L}t}\varrho(0)$. 
Moreover, we can define the quantum relative entropy of $\varrho_1$ with respect to $\varrho_2$ as:
\begin{equation}
    \mathcal{S}(\varrho_1\vert\varrho_2) \equiv \mbox{Tr}\left[\varrho_1\ln\varrho_1-\varrho_1\ln\varrho_2\right].
\end{equation}
It represents a  
distance measure between the states described by the two density matrices $\varrho_1$ and $\varrho_2$ \cite{vedral2002}.
We remember that  
Klein's inequality states that the quantum relative entropy $\mathcal{S}(\varrho_1\vert\varrho_2)$ is non-negative, and it is zero if and only if $\varrho_1=\varrho_2$ \cite{klein1931}.
We will use this result in the following development.
To better understand the structure of  Eq. (\ref{entroprodnew}), assuming that $\varrho=\varrho(t)$, we perform the following calculation: 
\noindent
\begin{equation}
\label{dere}
  \makebox[0pt][l]{\hspace*{-\mathindent}\(
    \begin{array}{rcl}
      &&  \displaystyle\left.-\frac{d}{d\tau}\mathcal{S}\left(e^{\mathcal{L}\tau}\varrho\vert\varrho_{eq}\right)\right\vert_{\tau=0}=\frac{d}{d\tau}\left[\mbox{Tr}\left(e^{\mathcal{L}\tau}\varrho\ln\varrho_{eq}\right)\left.-\mbox{Tr}\left(e^{\mathcal{L}\tau}\varrho\ln e^{\mathcal{L}\tau}\varrho\right)\right]\right\vert_{\tau=0}\\
 &&\displaystyle\quad=\left[\mbox{Tr}\left(e^{\mathcal{L}\tau}\mathcal{L}\varrho\ln\varrho_{eq}\right)-\mbox{Tr}\left(e^{\mathcal{L}\tau}\mathcal{L}\varrho\ln e^{\mathcal{L}\tau}\varrho\right)\left.-\mbox{Tr}\left(e^{\mathcal{L}\tau}\varrho\frac{d}{d\tau}\ln e^{\mathcal{L}\tau}\varrho\right)\right]\right\vert_{\tau=0},
  \end{array}
  \)}\hfill
\end{equation}
where the last term is zero since we know that $\mbox{Tr}\left(\varrho\frac{d}{dt}\ln\varrho\right)=0$ for any 
density matrix $\varrho$. Therefore, we obtain:
\begin{eqnarray}
\left.-\frac{d}{d\tau}\mathcal{S}\left(e^{\mathcal{L}\tau}\varrho\vert\varrho_{eq}\right)\right\vert_{\tau=0}=\mbox{Tr}\left(\mathcal{L}\varrho\ln\varrho_{eq}-\mathcal{L}\varrho\ln \varrho\right),
   \end{eqnarray}
which means that the rate of entropy production can be determined through the expression:
\begin{equation}
    \frac{d\mathcal{S}_p}{dt}=\left.-k_B\frac{d}{d\tau}\mathcal{S}\left(e^{\mathcal{L}\tau}\varrho\vert\varrho_{eq}\right)\right\vert_{\tau=0},
    \label{entrop}
\end{equation}
{  which has been first proved in Ref.\cite{spohn1978}.}
The latter can be further elaborated as follows:
\begin{eqnarray}
\nonumber
\frac{d\mathcal{S}_p}{dt}&=&\left.-k_B\lim_{\Delta\to 0}\frac{\mathcal{S}\left(e^{\mathcal{L}(\tau+\Delta)}\varrho\vert\varrho_{eq}\right)-\mathcal{S}\left(e^{\mathcal{L}\tau}\varrho\vert\varrho_{eq}\right)}{\Delta}\right\vert_{\tau=0}\\
\nonumber
&=&-k_B\lim_{\Delta\to 0}\frac{\mathcal{S}\left(e^{\mathcal{L}\Delta}\varrho\vert\varrho_{eq}\right)-\mathcal{S}\left(\varrho\vert\varrho_{eq}\right)}{\Delta}\\
&=&-k_B\lim_{\Delta\to 0}\frac{\mathcal{S}\left(e^{\mathcal{L}\Delta}\varrho\vert e^{\mathcal{L}\Delta}\varrho_{eq}\right)-\mathcal{S}\left(\varrho\vert\varrho_{eq}\right)}{\Delta},
\label{diffrel}
   \end{eqnarray}
where we used the property $e^{\mathcal{L}\Delta}\varrho_{eq}=\varrho_{eq}$, defining the equilibrium solution.

The result in Eq. (\ref{diffrel}) is significant since it expresses the entropic production as the difference between the relative entropies $\mathcal{S}\left(e^{\mathcal{L}\Delta}\varrho\vert e^{\mathcal{L}\Delta}\varrho_{eq}\right)$ and $\mathcal{S}\left(\varrho\vert\varrho_{eq}\right)$.
We recall an important property: the relative entropy decreases monotonically under the action of any completely positive, trace-preserving map $\Phi$ acting on the density matrix,
\begin{equation}
       \mathcal{S}(\Phi\varrho_1\vert\Phi\varrho_2)\le    \mathcal{S}(\varrho_1\vert\varrho_2).
\end{equation}
This inequality is called monotonicity of quantum relative entropy, and it was first proved by Lindblad \cite{lindblad1974,lindblad1975}.
Therefore, the rate of entropy production in Eq. (\ref{diffrel}) is non-negative if the operator $e^{\mathcal{L}\Delta}$ is completely positive and trace preserving. 

It is known that the equation describing a density matrix provides a completely positive and trace-preserving evolution if and only if it can be put into the so-called Lindblad form \cite{gorini1976,lindblad1976,pascazio2017}. 
Thus, if we assume that our equation generates a completely positive and trace-preserving evolution (we proved this point only for the harmonic oscillator), the entropy production rate will always be non-negative
\begin{equation}
\frac{d\mathcal{S}_p}{dt}=-k_B\lim_{\Delta\to 0}\frac{\mathcal{S}\left(e^{\mathcal{L}\Delta}\varrho\vert e^{\mathcal{L}\Delta}\varrho_{eq}\right)-\mathcal{S}\left(\varrho\vert\varrho_{eq}\right)}{\Delta}\ge 0.
\label{posprod}
\end{equation}
{  We note that the non-negativity of quantum entropy production has been established in the literature through a variety of approaches and methodologies, which nonetheless lead to conclusions consistent with the result obtained here \cite{spohn1978,lebowitz1978,trushechkin2017,trushechkin2019,ruelle2001,pillet2002,kosloff2019}.} 

To show a further connection with non-equilibrium thermodynamics, we also introduce the Helmholtz free energy $\mathcal{F}=\mathcal{E}-T\mathcal{S}$, and study its evolution. Recalling the  previously discussed relation $\mathcal{H}_0=-k_BT\ln\varrho_{eq}+k_BT\left(\ln\frac{1}{Z_{qu}}\right)I$, we obtain
\begin{equation}
    \mathcal{F}=k_BT\mbox{Tr}(\varrho\ln\varrho-\varrho\ln\varrho_{eq})-k_BT\ln Z_{qu}.
    \label{helm}
\end{equation}
It can be seen that at equilibrium the free energy takes on the asymptotic value $\mathcal{F}=-k_BT\ln Z_{qu}$, which corresponds to the standard expression of equilibrium statistical mechanics.
The first part of Eq. (\ref{helm}) can be identified with a relative entropy, and we can indeed write
\begin{equation}
    \mathcal{F}=k_BT\mathcal{S}(\varrho\vert\varrho_{eq})-k_BT\ln Z_{qu}.
\end{equation}
On the one hand, we can use the previously mentioned Klein's inequality $\mathcal{S}(\varrho\vert\varrho_{eq})\ge 0$, and immediately see that free energy always takes values larger than its equilibrium value
\begin{equation}
    \mathcal{F}\ge -k_BT\ln Z_{qu}.
\end{equation}
On the other hand, the free energy always decreases as time increases since
\begin{equation}
    \frac{d\mathcal{F}}{dt}=-\frac{1}{T}\frac{d\mathcal{S}_p}{dt}. 
\end{equation}
This underlines that a negative free energy change is a necessary condition for spontaneity (irreversibility). Altogether, our analysis confirms the complete consistency between the quantum formalism and macroscopic non-equilibrium thermodynamics.

\section{Numerical results}

In this final Section, we describe the implementation and the results obtained for the evolution equation of the density matrix for the harmonic oscillator.
The HO evolution equation was projected onto the energy basis, and solved numerically to observe the thermodynamics of the system during its relaxation towards thermal equilibrium.

\begin{figure*}[t!]
\centering
\includegraphics[width=15cm]{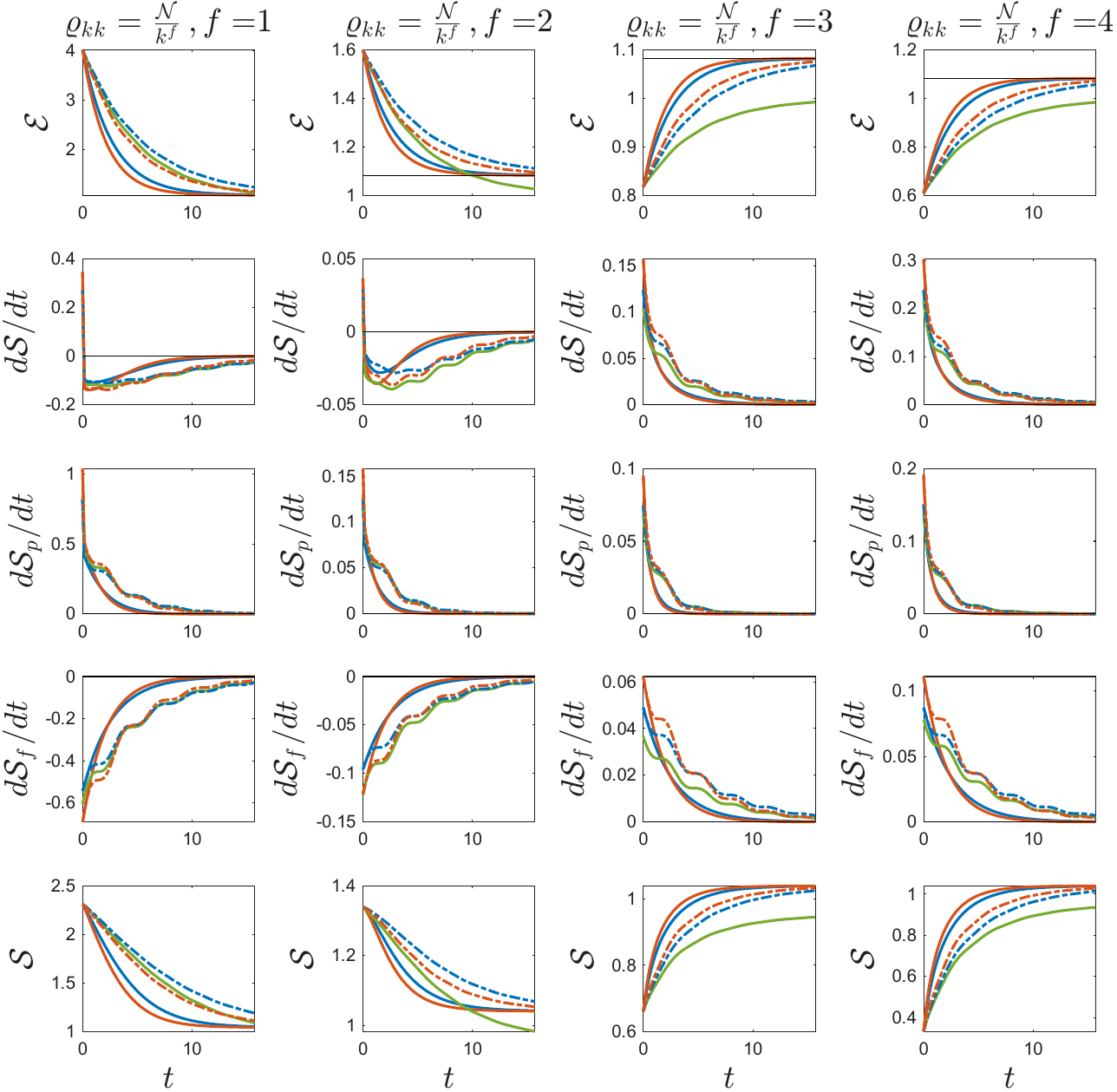}
\caption{\label{oscill}  {Results for the quantum harmonic oscillator in contact with a thermal bath. In each column are represented the plots corresponding to an initial density matrix (16$\times$16) given by $\varrho_{kk}(0)=\mathcal{N}/k^f$, with $f=1,2,3,4$ ($\mathcal{N}$ is a normalizing factor). In the first row, we show the total energy $\mathcal{E}$, which must converge to $\frac{1}{2}\hbar \omega+\hbar \omega/(e^{\frac{\hbar \omega}{k_B T}}-1)$. In the following rows, one can find the behavior of the total entropy rate $\frac{d\mathcal{S}}{dt}$, the entropy production rate $\frac{d\mathcal{S}_p}{dt}$, which is always positive, and the entropy flow rate $\frac{d\mathcal{S}_f}{dt}$. In the last row, we plot the entropy $\mathcal{S}$. Solid blue and red lines correspond to Hermitian and non-Hermitian friction operators with $\beta_p=m^2\omega^2\beta_q$. Dashed blue and red lines correspond to Hermitian and non-Hermitian friction operators with $\beta_q=0$. Solid green lines correspond to the Caldeira-Leggett model. We adopted the parameters $\hbar=1$, $k_B=1$, $T=1$, $\beta_p=0.2$, $\omega=1$, and $m=1$ in arbitrary units. We calculated 1000 times the exponential of a 256$\times$256 matrix with a time step $\pi/200$.}}
\end{figure*}

The first results are displayed in Fig. \ref{oscill}, where we plot the evolution of main quantities over time for different initial conditions, taking a 16x16 density matrix as a practical example of implementation.
In this Figure, one can compare five different models. The solid blue lines correspond to the model with Hermitian friction operators, with $\beta_p=m^2\omega^2\beta_q$ (ensuring complete positivity), see Eq. (\ref{har}). 
The solid red lines correspond to the model with non-Hermitian friction operators, with $\beta_p=m^2\omega^2\beta_q$ (ensuring complete positivity), see Eq. (\ref{harbis}).
The dashed blue and red lines correspond to models with Hermitian and non-Hermitian friction operators, respectively, when $\beta_q=0$. Notably, the blue dashed lines describe the model discussed in Ref. \cite{giordano2025}, and the red dashed lines the model introduced in Refs. \cite{oliveira2016,oliveira2023,oliveira2024}. In these cases, complete positivity is not guaranteed. 
Finally, the solid green lines correspond to the Caldeira-Leggett model \cite{caldeira1981,caldeira1983,caldeira1983bis}, where $\beta_q=0$ and $\Theta^p=p$. 
In this case, neither complete positivity nor thermodynamic consistency is guaranteed.

In Fig. \ref{oscill}, we considered as initial condition a mixed state composed of energy eigenfunctions with probability $p_k=\mathcal{N}/k^f$, for chosen values of the parameter $f$ ($\mathcal{N}$ is a normalizing factor, and $k=1$ corresponds to the fundamental state). 
The initial density matrix takes the form $\varrho_{kk}(0)=\mathcal{N}/k^f$, $\varrho_{kh}(0)=0$ if $h\neq k$.
It is observed that for large $f$ the probability distribution decays sharply with $k$, restricting the system to low-energy states and yielding a correspondingly low initial energy.
In contrast, for small $f$, the decay is slower, enabling substantial occupation of high-energy states and resulting in a higher initial energy. 
Recall that the asymptotic mean energy at thermal equilibrium is given by 
$\frac{1}{2}\hbar \omega+\hbar \omega/(e^{\frac{\hbar \omega}{k_B T}}-1)$.
In this construction, the selected superpositions of states are employed to define two initial conditions whose energies exceed the thermal equilibrium value, and two others whose energies lie below it. The first two cases (with $f=1,2$) are shown in the first two columns of Fig. \ref{oscill}, and the other two (with $f=3,4$) in the next two columns. 

In the panels of the first row, we observe the evolution of the internal energy. 
In the first two columns, the total energy decreases because of the outgoing heat, and in the other two, the energy increases because of the incoming heat. 
These heat flows generate an entropy flow rate that can be negative or positive.
This is evident in the panels of the fourth row of Fig. \ref{oscill}, where the entropy flow rate is negative in the first two cases and positive in the other two.  The entropy production rate is always positive, in accordance with the second law of thermodynamics, see the third row of Fig. \ref{oscill} and Eq. (\ref{posprod}).
This term, in fact, corresponds to the irreversibility and spontaneity of the process rather than the direction of heat flow.
The total entropy rate can also have a sign that depends on the initial conditions and the state of progress of the relaxation process (see second row in Fig. \ref{oscill}).  
In the last row of the Figure, we show the evolution of the total entropy $\mathcal{S}$, which may increase or decrease depending on thermal fluxes.

From these results, we conclude that the Caldeira-Leggett model does not comply with thermodynamics, as the asymptotic internal energy does not match to the value predicted by the canonical quantum distribution. 
The other models  appear to be consistent with thermodynamics, as they comply with all the evolutions predicted by the first and second laws, discussed in Section \ref{thermo}. Moreover, we checked that all eigenvalues of $\varrho$ remain positive during the time evolution of previous system. 

\begin{figure*}[t!]
\centering
\includegraphics[width=15cm]{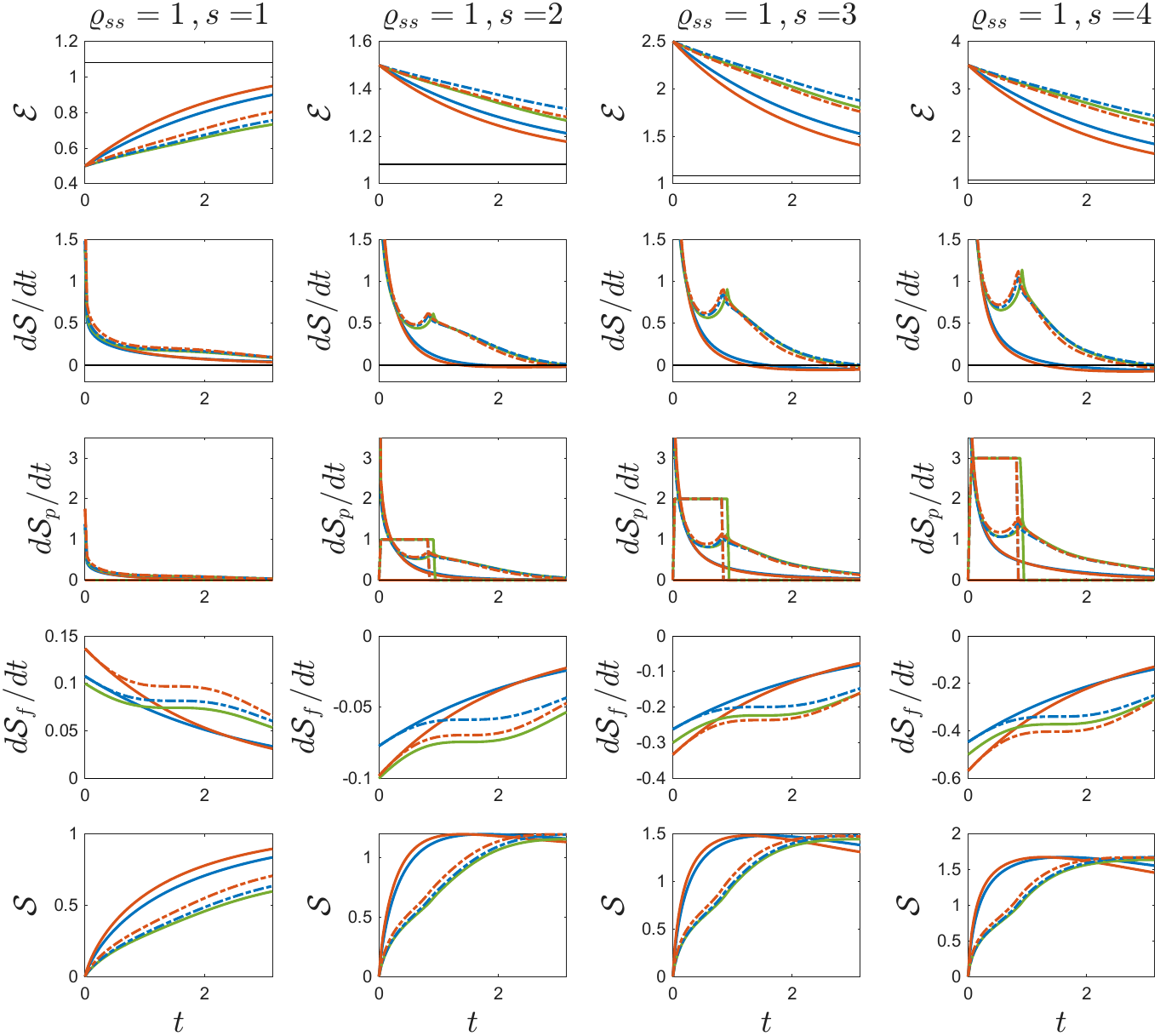}
    \caption{\label{oscillbis}  {Results for the quantum harmonic oscillator in contact with a thermal bath. In each column are represented the plots corresponding to an initial density matrix (16$\times$16) given by $\varrho_{ss}(0)=1$, with $s=1,2,3,4$ ($\mathcal{N}$ is a normalizing factor, $\varrho_{kk}(0)=0$ if $k\neq s$). In the first row, we show the total energy $\mathcal{E}$, which must converge to $\frac{1}{2}\hbar \omega+\hbar \omega/(e^{\frac{\hbar \omega}{k_B T}}-1)$. In the following rows, one can find the behavior of the total entropy rate $\frac{d\mathcal{S}}{dt}$, the entropy production rate $\frac{d\mathcal{S}_p}{dt}$, which is always positive, and the entropy flow rate $\frac{d\mathcal{S}_f}{dt}$. In the last row, we plot the entropy $\mathcal{S}$. In the third row, we also show the number of negative eigenvalues of $\varrho(t)=1$, revealing the positivity violation. Solid blue and red lines correspond to Hermitian and non-Hermitian friction operators with $\beta_p=m^2\omega^2\beta_q$. Dashed blue and red lines correspond to Hermitian and non-Hermitian friction operators with $\beta_q=0$. Solid green lines correspond to the Caldeira-Leggett model. We adopted the parameters $\hbar=1$, $k_B=1$, $T=1$, $\beta_p=0.2$, $\omega=1$, and $m=1$ in arbitrary units. We calculated 1000 times the exponential of a 256$\times$256 matrix with a time step $\pi/1000$.}}
\end{figure*}

Nevertheless, these observations correspond to initial conditions specified by particular mixed states. If we instead consider pure initial states, we obtain the results shown in Fig. \ref{oscillbis}, which reveal other interesting phenomena.
Here, we consider the four initial conditions corresponding to the pure states characterized by the first four energy states of the harmonic oscillator. This means that we consider the initial density matrix taking the form $\varrho_{ss}(0)=1$ for $s=1,2,3$ or 4, $\varrho_{kk}(0)=0$ if $k\neq s$, and $\varrho_{kh}(0)=0$ if $h\neq k$.
These four initial conditions lead to the evolutions shown in Fig. \ref{oscillbis}, where we exhibit the same physical quantities considered in Fig. \ref{oscill} for a short interval of time.
The evolution of energy shows no significant changes, and even in this case, only the Caldeira-Leggett model does not respect the correct energetic thermodynamic asymptotic value.
In the entropy rate and entropy production rate plots (second and third rows), we observe (in the last three columns) the presence of small peaks that need to be understood. 
To this end, in the entropy production plots, we have also superimposed the time evolution of the number of negative eigenvalues of the density matrix. 
We see that the three models with $\beta_q=0$ (with Hermitian and non-Hermitian friction operators, and the Caldeira-Leggett model) have some unphysical negative eigenvalues in the first time interval, showing the non complete positivity of these models, at least for these initial conditions. This is consistent with what was demonstrated previously: the evolution equation can be of the Lindblad type only if   $\beta_p>0$ and $\beta_q>0$.
Once again, we see that it is essential to include the friction and noise terms in the second classical Hamilton equation, Eq. (\ref{hambis}), in order for the canonical quantization to work correctly. 
The peaks observed in the three models with $\beta_q=0$ are therefore artifacts of the lack of positivity of the density matrix, and are in fact absent in the two models with $\beta_q>0$ (solid blue and red lines). 

We therefore conclude that the two models defined in Eqs. (\ref{har}) and (\ref{harbis}) -- with $\beta_p>0$, $\beta_q>0$, and Eq. (\ref{inte1}) or (\ref{inte2}) satisfied -- are correct from the thermodynamic point of view (positive entropy production) and also from the mathematical point of view (complete positivity).
Moreover, these two properties are closely related to each other through the monotonicity of quantum relative entropy, as discussed in Section \ref{thermo}.
Should one want to distinguish between Eq. (\ref{har}) and Eq. (\ref{harbis}), it may be said that Eq. (\ref{har}) is preferable because it adopts Hermitian friction operators and they can therefore be associated with real physical observables. 
An important issue that remains open at present is the verification of the complete positivity of the general models defined in Eqs. (\ref{maindiffforces}) and (\ref{maindiffforcesbis}) for an arbitrary potential acting on the quantum system, and not only for the harmonic oscillator. 

\section{Conclusions}

In this work we have developed a unified classical-to-quantum framework for the stochastic description of open systems, grounded in a generalized Langevin dynamics where friction and noise act symmetrically on both Hamilton equations. 
At the classical level, this structure leads to a generalized Klein–Kramers equation that preserves the canonical Gibbs distribution and is fully consistent with the first and second laws of thermodynamics. 
This result substantiates the physical relevance of including both dissipative and stochastic contributions in the equations for position and momentum, revealing a minimally required symmetry for thermodynamic consistency.

By applying canonical quantization to the classical Klein–Kramers operator, we derived two distinct quantum master equations, corresponding respectively to Hermitian and non-Hermitian friction operators. 
Despite their structural differences, both formulations yield Lindblad-type dynamics for the harmonic oscillator under the same universal constraint on the friction coefficients. 
Remarkably, this constraint is independent of the choice of operator representation, suggesting that complete positivity in the quantum regime consistently mirrors the classical requirement of symmetric dissipation and noise. This generality strengthens the conceptual bridge between the classical and quantum descriptions of nonequilibrium statistical mechanics.
Moreover, it is interesting to note that the quantum optical master equation used in countless applications can be expressed in the form predicted by our approach.

Our results demonstrate that friction and noise must be included in both Hamilton equations to guarantee a fully thermodynamically consistent and completely positive quantum evolution. 
The formalism naturally provides generalized quantum expressions for heat, work, and entropy production, leading to compact formulations of the quantum first and second laws, closely connected with the monotonicity of quantum relative entropy.

The methodology introduced here can be readily adapted to a broad class of nanoscale and mesoscopic systems of contemporary interest, ranging from quantum dots to optomechanical platforms. It thus offers a robust and physically transparent foundation for exploring quantum thermodynamics in regimes where stochasticity, dissipation, and quantum coherence interplay in essential ways.

A natural perspective for future research is to establish that the symmetry of friction and noise in the Hamilton equations, for arbitrary potential energy landscapes, systematically leads—through canonical quantization—to master equations of Lindblad type, and therefore to completely positive quantum evolutions. This question can be approached by analyzing additional simple models for which the energy basis allows fully analytic expressions for all operators, such as the infinite potential well, and then extending the analysis to general potential energies. Depending on the complexity of the potential, this program may be pursued analytically or numerically.
{  Another promising line of investigation concerns the generalization of the present models to non-Markovian dynamics, in which memory effects become relevant.}

\vspace{1cm}

\ack{G.F. and G.P. have been supported by GNFM - Gruppo Nazionale per la Fisica Matematica(INdAM), Italy. G.F. and G.P.’s
research is funded by the European Union (EU) - Next Generation EU. G.P. and G.F. are supported by PNRR, Italy, National Center
for HPC, Italy, Big Data and Quantum Computing - M4C2 - I 1.4, Italy (grant number N00000013, CUP D93C22000430001) - Spoke
5 (Environment and Natural Disasters). G.P. and C.B. are supported by the Project of National Relevance (PRIN), Italy, financed by
EU - Next- GenerationEU - NRRP - M4C2 - I 1.1, CALL PRIN 2022 PNRR (Project P2022KHFNB, CUP D53D23018910001) granted by
the Italian MUR. G.P. is supported by the PRIN, Italy, financed by Eu - Next- Generation EU - NRRP - M4C2 - I 1.1, CALL PRIN 2022
(Project 2022XLBLRX, CUP D53D23006020006) granted by the Italian MUR. G.F. is supported by the PRIN, Italy, financed by EU -
Next-Generation EU - NRRP - M4C2 - I 1.1, CALL PRIN 2022 PNRR (Project P2022MXCJ2, CUP D53D23018940001) and CALL PRIN
2022 (Project 2022MKB7MM, CUP D53D23005900006) granted by the Italian MUR. G.F. is also supported by ‘‘Istituto Nazionale
di Fisica Nucleare’’ (INFN), Italy, through the project QUANTUM. S.G. has been supported by ‘‘Central Lille, France’’ and
‘‘Région Hauts de France’’ under the project StaMeNa (Statistical mechanics for macromolecular structures of nanotechnology).}

\vspace{1cm}


\end{document}